\documentclass{JFM-FLM_Au}

\usepackage{multirow}
\usepackage{subfigure}
\usepackage{color}
\usepackage{float}
\usepackage{booktabs} % Added for better table formatting

\lefttitle{A. Yakeno et al.}
\righttitle{Journal of Fluid Mechanics}

\title{DMR effect on drag reduction of a streamlined body measured by Magnetic Suspension and Balance System}

\author{A. Yakeno, H. Okuizumi, K. Inokuma, Y. Watanabe}

\affiliation{Institute of Fluid Science, Tohoku University, Katahira 2-1-1, Aoba-ku, Sendai, Miyagi, 980-8577, Japan}

\corresau{A. Yakeno, \email{aiko.yakeno@tohoku.ac.jp}}

\begin{document}
\maketitle

\begin{abstract}
This study experimentally investigates the aerodynamic drag reduction capabilities of distributed micro-roughness (DMR) coatings on a streamlined model, utilising the 1-m magnetic suspension and balance system (MSBS) at Tohoku University. Previous direct numerical simulations (DNS) indicated that DMR can mitigate turbulent-energy growth by suppressing Tollmien--Schlichting (TS) waves and influencing the breakdown of streamwise vortices. The present work provides the first experimental validation of these effects using an interference-free MSBS, which is essential for accurate measurement in the laminar and transitional regimes. A streamlined model was tested with two rows of artificial tripping tape to induce transition; the DMR height was approximately 1\% of the local boundary layer thickness, significantly smaller than typical roughness elements. Direct aerodynamic drag measurements using the MSBS revealed a substantial reduction of up to 43.6\% within the transitional flow regime. Crucially, integrated analysis using wall-resolved large-eddy simulations (LES) and dynamic oil-flow visualisation confirmed that this benefit does not mainly originate from the suppression of flow separation. The LES drag decomposition established that the total pressure-drag budget is subordinate to skin friction, a finding complemented by oil-flow observations, which revealed qualitatively similar flow patterns regardless of the surface condition. Consequently, the observed drag reduction is primarily ascribed to friction drag reduction achieved through the modification of the boundary layer state. These findings provide compelling experimental evidence for the efficacy of DMR and offer valuable insights for optimising surface designs for passive flow control.
\end{abstract}


%============================================================================
% 							Introduction
%============================================================================
\section{Background}

Laminarisation has long been projected as a means to significantly diminish friction drag over aerodynamic surfaces. Since the 1940s, diverse wing geometries have been proposed to retard the boundary layer transition from laminar to turbulent flow. In Japan, Ichiro Tani made substantial contributions to laminar wing development; however, contemporary reports \citep{Tani1935, Tani1940} indicated that the full benefits of laminarisation were not realised at the time due to the considerable surface roughness inherent in aircraft manufacturing. For many decades, conventional wisdom posited that an exceptionally smooth surface was essential to prolong transition delay.

Nevertheless, research in this area expanded significantly from the 1990s onward, with groups like Saric's initiating extensive investigations into surface roughness, including flight tests \citep{Saric1997}. They notably demonstrated that Discrete Roughness Elements (DREs) could delay transition in three-dimensional boundary layers by suppressing crossflow instability. Two unstable modes are recognised within crossflow instability: a stationary mode and a traveling mode. It is increasingly evident that the transition route is not solely dependent on turbulence intensity, as previously understood \citep{Morkovin1969, Saric1994}. Instead, it involves a more intricate process influenced by the disturbance frequency band and incorporating theoretically predicted linear mode interference \citep{Schaffarczyk2017, Yakeno2021, nakagawa2023effects, Mori2024}. Consequently, the transition mechanisms around wings under genuine flight conditions remain an active area of investigation \citep{Mori2024IUTAM, Mori2024AIAA}. The development of flow control strategies for drag reduction is a continuous and multifaceted field, encompassing a wide range of methods tailored to different flow regimes. We contend that experimental demonstrations of effectiveness remain crucial, complementing idealised numerical simulations.

Despite growing momentum to verify drag reduction performance through testing, achieving extensive laminarisation in large aircraft remains a formidable challenge. Initiatives like the Airbus Breakthrough Laminar Aircraft Demonstrator in Europe (BLADE) project \cite{AirbusBLADE} underscore the complexities involved. Challenges associated with laminarisation in large aircraft primarily stem from their considerable size, which typically leads to the boundary layer transition occurring relatively upstream, resulting in a larger proportion of turbulent boundary layer. Additionally, the swept-back angle of main wings induces complex three-dimensional flow distributions.

While roughness is often considered detrimental to laminar flow, certain configurations have shown promise. For instance, in hypersonic boundary layers, the transition delay effects of porous surfaces \citep{Rasheed2002Experiments, Fedorov2003Stabilization, Lukashevich2018Passive, Lim2022Simulation, Lim2023Turbulent, Running2023Attenuation} and finite-height surface roughness \citep{Holloway1964Effect, Duan2012Stabilization, Fong2015Second} have been actively studied. This transition delay effect in hypersonic flows merits attention not only for friction drag reduction but also for thermal protection. Regarding porous surfaces, \cite{Malmuth1998Problems} proposed that an ultrasonically absorbing porous surface can stabilise the acoustic-wave-like Mack second mode. This concept was initially examined in theoretical studies using inviscid stability analysis. Later, \cite{Fedorov2001Stabilization} utilised viscous linear stability analysis (LST) to show that porous surfaces can effectively stabilise the Mack second mode in hypersonic laminar boundary layers. This theoretical concept was experimentally verified by \cite{Rasheed2002Experiments} in a hypersonic shock tunnel. More recently, \cite{Li2024Porous} have attempted to optimise porous shapes using LST theory.

However, whilst the attenuation of turbulence fluctuations has been extensively discussed in previous hypersonic flow studies, that phenomenon and the resulting drag reduction effect in subsonic flows has been significantly limited. It is in this context that our research on Distributed Micro-Roughness (DMR) makes a unique contribution. Recently, \cite{hamada2023drag} employed Direct Numerical Simulations (DNS) to demonstrate the drag reduction efficacy of DMR on transitional flows governed by Tollmien-Schlichting (T-S) instability, a quintessential transition mechanism. DMR, characterised by randomly distributed micron-sized roughness elements resembling a sandy surface, offers distinct advantages over conventional DREs. Specifically, DMR exhibits a more robust effect across varying flow directions and is inherently simpler to implement. The drag-reducing potential of DMR was initially inspired by Tani's foundational reference \citep{Tani1989} and the seminal experiments conducted by Kohama's group \citep{Oguri1996, Oguri1998, Kikuchi2004}. Generally, such random surface roughness has not been extensively investigated for its friction-drag-reduction effect in subsonic flows \citep{Nikuradse1933, Schlichting1936, Hama1954, Perry1969, Bhaganagar2004, Flack2010, Busse2017, chung2021predicting}; it is more commonly deployed as an artificial disturbance to promote transition in wind tunnel experiments aimed at replicating realistic flight conditions of a large scale. In this context, our confirmation of the drag reduction effect of DMR in the subsonic region \citep{hamada2023drag, patent7609489} was somewhat groundbreaking.

The accurate measurement of skin friction drag has been a fundamental challenge in experimental fluid dynamics, and a key reason for the limited experimental discussion on drag reduction is the inherent difficulty in precisely quantifying these changes. Traditional methods exist to calculate the friction coefficient, such as fitting boundary layer distributions measured by hot-wire anemometry or PIV, or employing the Clauser method \citep{butt2018transition}. However, precisely defining the wall position, especially on rough surfaces, remains a significant challenge for these indirect approaches. To overcome these measurement difficulties, various direct experimental techniques have been developed. For instance, Kohama's group measured drag changes by suspending a flat plate in a wind tunnel using a piano wire and a load cell \citep{Oguri1996}. At the University of Melbourne, measurements are performed using a floating element shear stress meter, which translates the displacement of a large floating plate into a force measurement using load cells \citep{squire2016comparison, chung2021predicting}. More recently, Mochizuki's group has devised an advanced instrument that calculates shear stress by installing a very small floating element on a section of the wall and precisely detecting its minute displacement, caused by fluid shear stress, using strain gauges, optical sensors, or optical fiber sensors \citep{nonomiya2024development}. These advanced direct measurement methods offer a higher degree of reliability.

Concurrently, our research institute has been developing a unique and powerful tool for aerodynamic measurement: the Magnetic Suspension and Balance System (MSBS). Accurate aerodynamic drag measurements are crucial for advancing aerospace engineering, from designing next-generation aircraft to optimising high-speed vehicles. While traditional wind tunnel testing is indispensable, it often faces a fundamental challenge: the interference caused by mechanical support systems. These physical mounts, essential for model stability, introduce undesirable aerodynamic disturbances, obscure critical flow features, and can significantly alter measured forces and moments \citep{Pankhurst1952, joppa1973wind, britcher2023wind}. This inherent limitation complicates the accurate characterisation of aerodynamic phenomena and the validation of computational fluid dynamics (CFD) simulations.

To address these challenges, the MSBS emerged as a revolutionary technology in experimental aerodynamics. Conceived in the mid-20th century, the MSBS levitates a wind tunnel model in an airstream using precisely controlled electromagnetic forces, thereby entirely eliminating mechanical support interference. This unique capability allows for truly "free-flight" conditions within the wind tunnel environment, offering unprecedented fidelity in aerodynamic force and moment measurements, as well as providing unobstructed optical access for flow visualisation techniques \citep{tuttle1983magnetic, lawing1987magnetic, garbutt1992propulsion}. Early MSBS development, notably by NASA and various academic institutions across the United States, Europe, and Japan, focused on establishing the fundamental principles of stable magnetic levitation and control for aerodynamic models. The progress of MSBS technology in Japan, particularly at the National Aerospace Laboratory (NAL, now part of JAXA), is detailed by Sawada et al. \citep{sawada1995recent}. These pioneering efforts demonstrated the feasibility of suspending models with multiple degrees of freedom, paving the way for detailed studies of complex flow phenomena without the confounding effects of sting supports. Over the decades, advancements in electromagnet design, power electronics, digital control systems, and computational capabilities have progressively enhanced the performance, stability, and versatility of MSBS facilities. This evolution has enabled the testing of larger models, at higher dynamic pressures, and with greater precision, making MSBS an invaluable tool for both fundamental research and applied aerodynamic development.

The Institute of Fluid Science at Tohoku University developed a 1-meter MSBS, one of the largest in the world \citep{okuizumi2018sports}. Such large-scale testing equipment can minimize measurement errors. Despite significant progress, the design and operation of an MSBS facility remain a complex engineering challenge, demanding sophisticated integration of magnetic fields, control algorithms, and aerodynamic measurement techniques. However, the unparalleled advantages of interference-free testing continue to drive innovation in this field. The development and refinement of MSBS technology are crucial for pushing the boundaries of aerodynamic understanding and validating the performance of advanced aerospace designs under highly realistic flow conditions. Our 1-m MSBS has primarily been utilised for measuring the aerodynamic forces of blunt objects \citep{okuizumi2024aerodynamic, okuizumi2025wind}. This study aims to leverage the interference-free capability of the MSBS to accurately evaluate the effect of DMR on friction drag by measuring the total aerodynamic forces of a streamlined model where flow separation is suppressed.

To address these objectives, the remainder of this paper is structured as follows. Section \ref{sec:method} provides a detailed overview of the experimental setup, encompassing the 1-m MSBS facility, the streamlined model design, and the specifics of the DMR coating application and tripping tape configurations. This section also outlines the methodology employed for aerodynamic force measurements. Section \ref{sec:measurement} then presents the validation of measurements, details on measurement error, and the stability of the model during testing. Subsequently, Section \ref{sec:results} presents the experimental results, including comparisons with Large Eddy Simulations. It commences with a comparison of the drag characteristics between the Plane and DMR-coated models across different experimental phases, followed by a quantitative analysis of the observed drag reduction percentages. This section also discusses the inferred critical Reynolds numbers for each test condition. Finally, Section \ref{sec:conclusion} summarises the key conclusions drawn from this research and outlines potential avenues for future work.

%============================================================================
% 					        Method
%============================================================================
\section{Method\label{sec:method}}

\subsection{Wind Tunnel Facility}
The experiments were performed in the low-turbulence wind tunnel at the Institute of Fluid Science, Tohoku University. This facility has been extensively utilised for numerous transition studies owing to its outstanding characteristics \citep{kohama1982performance, kohama1987some, Ito_Kobayashi_Kohama_1992, kohama1992tohoku, kohama1994traveling, kohama1999effective, suzuki2024experimental}. It boasts exceptionally low turbulence intensity, a critical feature for fundamental fluid dynamics research, alongside precise control over airflow conditions. This combination enables the generation of highly stable and uniform flows, rendering it ideally suited for detailed investigations into boundary layer phenomena and heat transfer.

The wind tunnel features two primary test section configurations. The open-jet test section possesses an octagonal cross-section with a diagonal length of $0.81\,\text{m}$. When connected with the 1-m MSBS, or in its closed-wall configuration, the test section also exhibits an octagonal shape, with a distance across flats of $1.01\,\text{m}$. Furthermore, the facility maintains remarkable turbulence intensity. Within the 1-m MSBS test section, the turbulence intensity is less than $0.06\,\%$ for free-stream velocities ranging from $5\,\text{m}\,\text{s}^{-1}$ to $50\,\text{m}\,\text{s}^{-1}$. For the closed-wall test section, it is less than $0.04\,\%$ for free-stream velocities ranging from $5\,\text{m}\,\text{s}^{-1}$ to $70\,\text{m}\,\text{s}^{-1}$, thereby ensuring high-quality experimental conditions. The facility is also comprehensively equipped with various measurement systems for the accurate capture of flow velocity, temperature, and pressure.

\begin{figure}[H]
\centering
\includegraphics[width=0.8\textwidth]{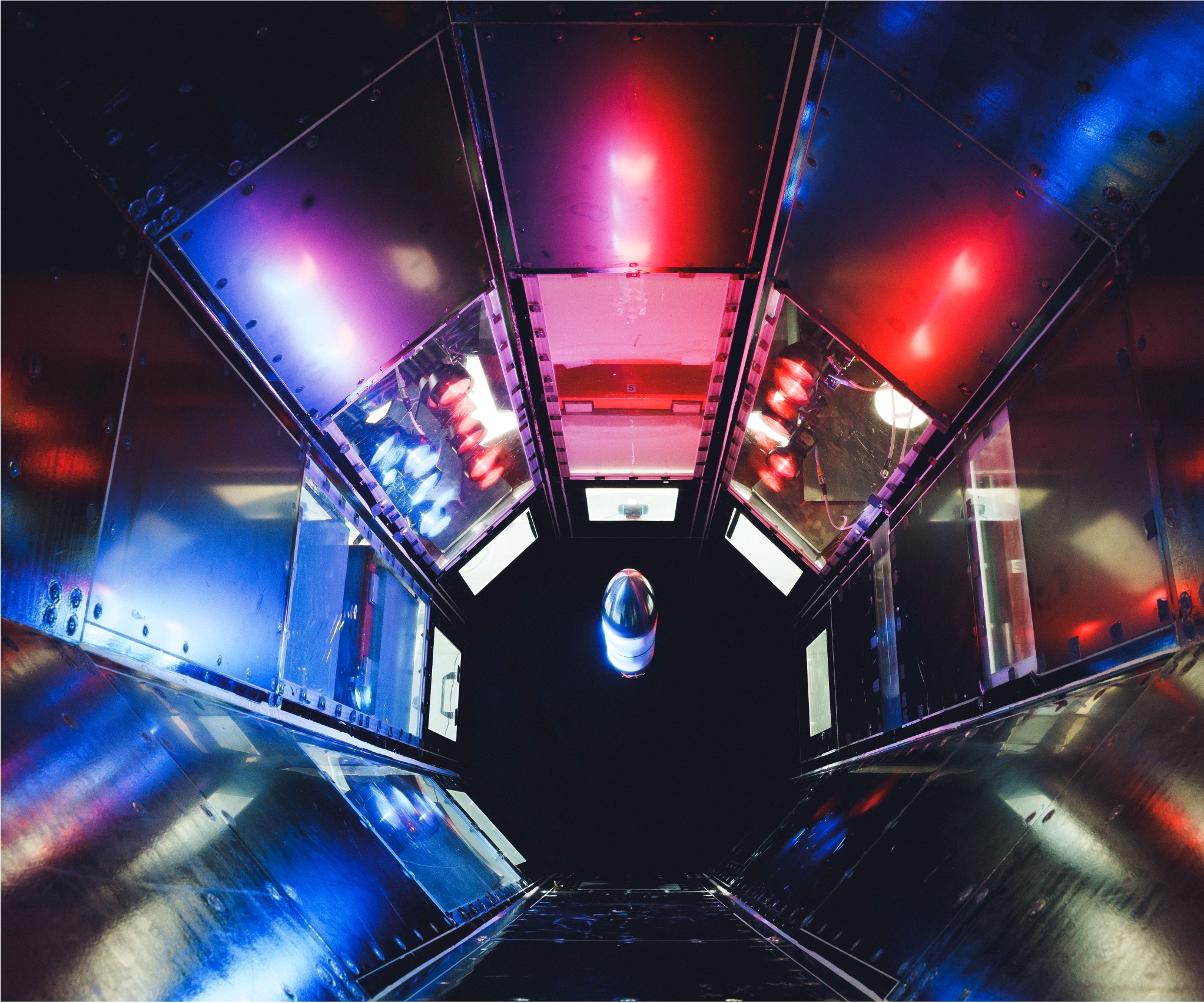}
\caption{The streamlined model in 1-m Magnetic Suspension and Balance System (1-m MSBS) installed in the low-turbulence wind tunnel. The system uses electromagnetic forces to suspend the test model without physical supports, allowing for highly accurate, interference-free aerodynamic measurements.}
\label{fig:msbs}
\end{figure}

\subsection{1-m Magnetic Suspension and Balance System (1-m MSBS)}

The experimental measurements in this study were conducted using the 1-m Magnetic Suspension and Balance System (MSBS) at the Institute of Fluid Science, Tohoku University. This facility is a large-scale, cutting-edge wind tunnel system that employs electromagnetic forces to suspend models within the test section without physical supports. This unique capability entirely eliminates interference from model supports, a common source of error in conventional wind tunnel testing. As one of the largest of its kind globally, the 1-m MSBS provides an unparalleled environment for highly accurate aerodynamic force measurements under more realistic flow conditions. A streamlined model in the 1-m MSBS installed in the low-turbulence wind tunnel is shown in Figure \ref{fig:msbs}.

The system is designed to precisely control the model's position and attitude within the airflow, offering six degrees of freedom (three translations and three rotations) via a sophisticated array of electromagnets. This precise control allows for detailed investigations of aerodynamic phenomena, including lift, drag, and moments. In the present experiment, the roll angle during levitation naturally settles at an angle determined by the model's slight centre-of-gravity offset. Consequently, only five-axis control was activated. This approach was justified as negligible roll rotation was observed during the actual measurements. Whilst the facility has been extensively utilised for measuring aerodynamic forces on blunt objects, its interference-free nature makes it exceptionally well-suited for the current study, enabling us to accurately evaluate the subtle effects of DMR on friction drag in a streamlined model without the confounding influence of support interference. The detailed methodology for the magnetic force control and measurement procedure is described in Appendix \ref{sec:appendix}.

\begin{figure}[H]
\centering
\vspace{2em}
\includegraphics[width=\textwidth]{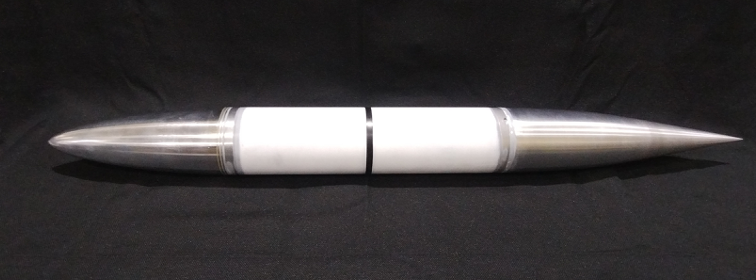}
\caption{A photograph of the streamlined test model.}
\label{fig:testmodel_photo}
\end{figure}

\subsection{Test Model Description}

A photograph of the streamlined model utilised in the test is shown in Figure \ref{fig:testmodel_photo}. The test model featured a central cylindrical section, measuring $0.40\,\text{m}$ in length and $0.10\,\text{m}$ in diameter. This cylinder was flanked by distinct leading and trailing edge sections, the profiles of which were derived from the NLF2-0415 laminar flow aerofoil \citep{somers1980design} and adapted to fit the $0.10\,\text{m}$ diameter constraint. The leading edge was smoothly faired from the nose to the $0.1\,\text{m}$ maximum diameter based on the NLF2-0415 forward profile. For the trailing edge, the aft profile of the NLF2-0415 was axially extended (stretched) by a factor of two to achieve the target overall model length (approximately $1.0\,\text{m}$) and to minimise flow separation at the rear. The overall design corresponds to a streamlined body with cusped nose and tail. However, due to manufacturing constraints and handling considerations, the tips are not perfectly sharp but are slightly rounded, a necessity to ensure structural integrity and prevent physical damage. Detailed digital microscope imaging confirms that the tips are smoothly rounded, not truncated (cut flat), with the rounding radius estimated to be of a small order. The trailing edge section, measuring approximately $0.406\,\text{m}$, was intentionally extended beyond the leading edge, which measured approximately $0.263\,\text{m}$. The overall length of the assembled model was therefore approximately $1.069\,\text{m}$.

\begin{figure}[H]
\centering
\subfigure[Leading and trailing edge sections]{
\includegraphics[width=1.0\textwidth]{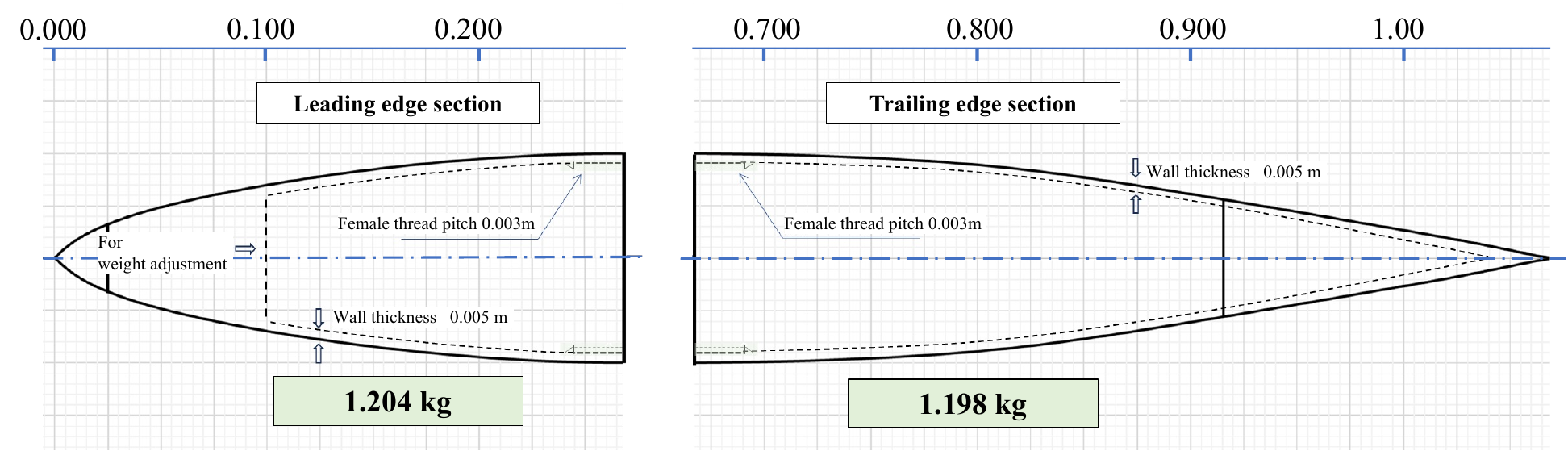}}
\subfigure[Cylindrical section, neodymium magnet, and two magnet-fixing blocks]{
\includegraphics[width=1.0\textwidth]{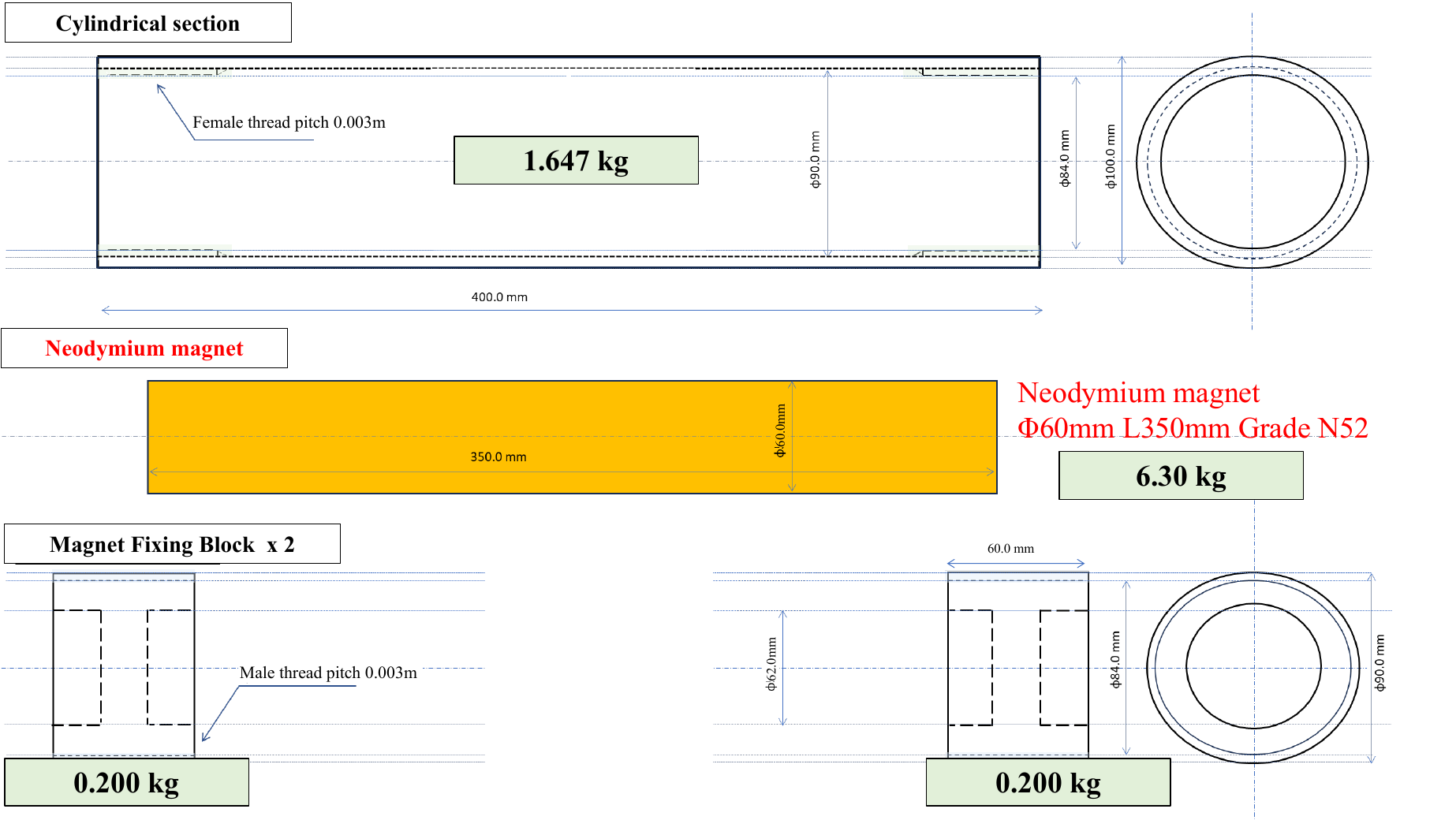}}
\caption{Geometry of parts of the streamlined test model.}
\label{fig:testmodel_parts}
\end{figure}

An overview of the model's constituent components is presented in Figure \ref{fig:testmodel_parts}: (a) depicts the removable leading and trailing edge sections, while (b) illustrates the central cylindrical section, the embedded neodymium magnet, and the two magnet-fixing blocks. The leading and trailing edge sections are secured to the main body using resin male threaded fasteners, which facilitates the installation of the internal magnet within the cylindrical section.

The model was primarily fabricated from machined aluminium, with all joints meticulously polished and filled with putty to eliminate gross surface discontinuities. Additional attention was paid to the joints in the removable sections that were unavoidable due to model construction, particularly the vertical seams shown in Figure \ref{fig:testmodel_parts}(a). These joints were carefully polished and left intact throughout the test campaigns. Detailed measurements using a laser microscope revealed a very small positive step height of approximately $1\text{-}10\,\mu\text{m}$ ($0.001\text{-}0.010\,\text{mm}$) at the junction. The location of the joint was intentionally chosen to minimise its impact on the measured drag. Given that the flow over the leading edge is an acceleration region, this minute step is unlikely to amplify turbulence. The boundary layer thickens significantly in the aft (tail) section, making this discontinuity in surface finish relatively minimal and gradual. In the last section, we present the results of oil flow visualisations, which show that the oil trapped at this junction moves slowly at low Reynolds number. While the effect of this small step on the separation/reattachment behaviour at lower speeds cannot be entirely ruled out, the same tail model was consistently used across all measurements. Importantly, all comparative measurements were performed under the same geometric conditions described above.

The embedded magnet is an N52-grade neodymium magnet, with a diameter of $0.060\,\text{m}$ and a length of $0.350\,\text{m}$, weighing approximately $6.3\,\text{kg}$. To ensure static balance of the model, the internal aluminium of both the leading and trailing edge sections was precisely machined such that each weighed approximately $1.2\,\text{kg}$. A $0.003\,\text{m}$ wide female threaded groove is machined into the inner surface of the central cylindrical section, where the magnet is housed. Following the insertion of the magnets, two resin blocks with male threads are attached to firmly secure them in position. The total mass of the assembled model is approximately $10.75\,\text{kg}$.

\begin{figure}[H]
\centering
\vspace{2em}
\includegraphics[width=1.0\textwidth]{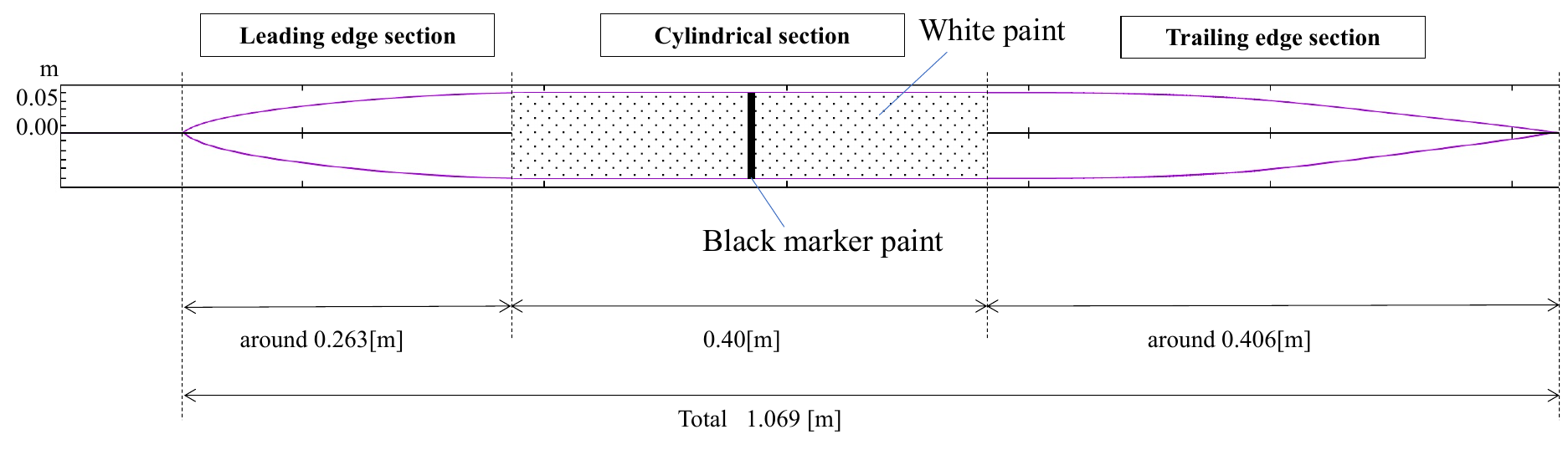}
\caption{Colouring of the streamlined test model as a side view cross-section.}
\label{fig:testmodel}
\end{figure}

For the optical sensing system, the central cylindrical section of the model was painted white, with $0.010\,\text{m}$ wide black marker lines, as shown in Figure \ref{fig:testmodel}. These were carefully applied to ensure a uniform surface finish. Following the application of the white and black paint to the cylindrical section, the entire model was polished with sandpaper up to 20,000 grit, achieving a smooth reference surface comparable to that of the aluminium. For the DMR-coated condition, only the leading and trailing edge sections of the model underwent this fine polishing. Prior to each measurement, the model surface was thoroughly wiped with isopropyl alcohol to remove any contaminants.

\begin{figure}[H]
\centering
\vspace{2em}
\subfigure[Tape 1]{
\includegraphics[width=0.45\textwidth]{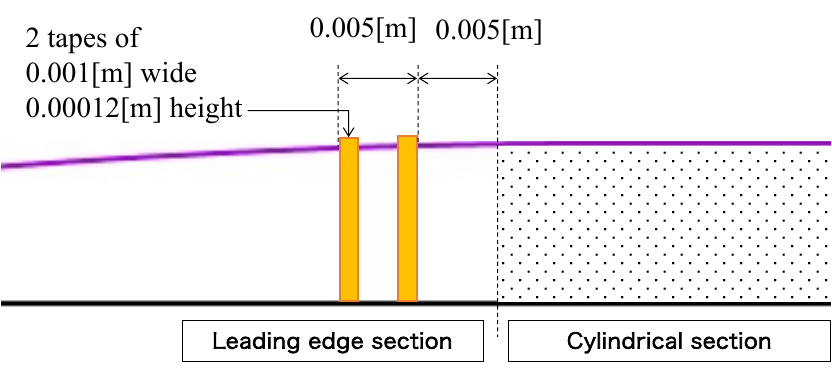}}
\subfigure[Tape 2]{
\includegraphics[width=0.45\textwidth]{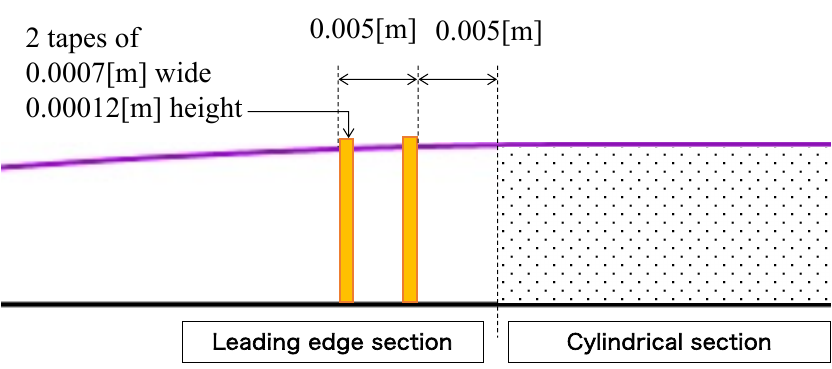}}
\caption{Trip tapes.}
\label{fig:tape}
\end{figure}

In the present study, given the inherent challenges in inducing boundary layer transition within a low-turbulence wind tunnel environment, two rows of trip tape were applied to the leading edge of the model. The efficacy of employing such trip tapes to promote transition is well-established in experimental fluid dynamics \citep{saric1991boundary, saric2002boundary}. As illustrated in Figure \ref{fig:tape}, a minor variation in tape width existed between the two distinct measurement series (Phase I and Phase II). Specifically, whilst the height and installation position of the tapes remained consistent at $0.00012\,\text{m}$ and $0.005\,\text{m}$, respectively, the tape width was $0.001\,\text{m}$ for the initial measurement series (Phase I) and $0.0007\,\text{m}$ for the subsequent series (Phase II).

Results from the Large Eddy Simulation (LES), detailed in a later section, indicate the boundary layer thickness upstream of the tape location under various free-stream velocities. These were approximately $0.00359\,\text{m}$ at a low speed of $10\,\text{m/s}$, $0.00240\,\text{m}$ at a medium speed of $25\,\text{m/s}$, and $0.00155\,\text{m}$ at a high speed of $50\,\text{m/s}$. Whilst the specific spacing of the trip tapes utilised in this investigation may not always precisely correspond to that which would induce the maximum amplification of Tollmien-Schlichting (T-S) wave energy, it was nonetheless sufficient to provoke a measurable increase in drag attributable to transition across the entire range of velocities examined. This unequivocally confirms the effectiveness of the applied tripping tape in robustly inducing boundary layer transition for the objectives of our aerodynamic measurements.

\subsection{DMR Coating \label{sec:dmr}}
\subsubsection{Phase I}
The effectiveness of the DMR concept was initially evaluated through experiments conducted in Phase I. In this initial setup, the DMR primarily consisted of glass beads affixed with a transparent adhesive. These roughness elements were Fuji Glass Beads (model number FGB-320) manufactured by Fuji Seisakusho, with diameters ranging from $38$ to $53\,\mu\text{m}$.

\begin{figure}[H]
\centering
\subfigure[Glass beads adhered to a sample]{
\includegraphics[width=0.4\textwidth]{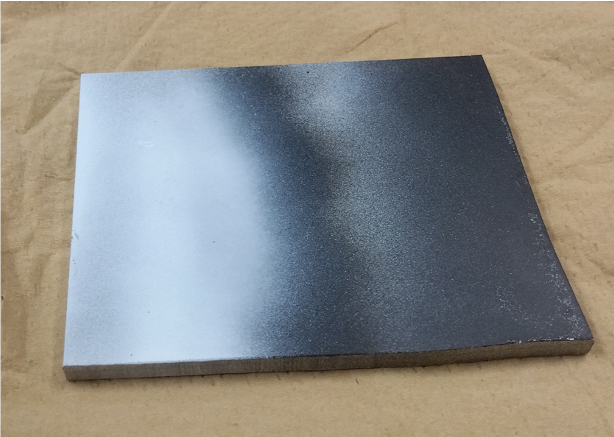}}
\hspace{2em}
\subfigure[Magnified image of the sample surface]{
\includegraphics[width=0.45\textwidth]{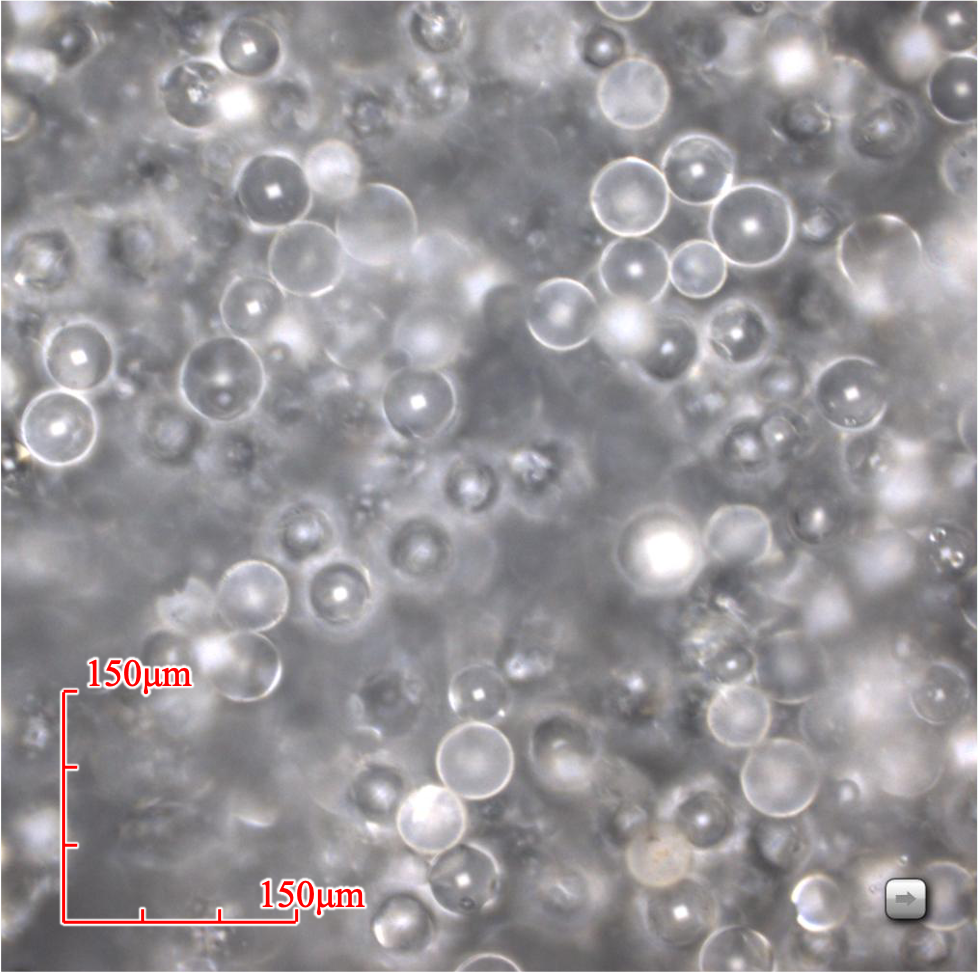}}
\caption{Characteristics of the glass-DMR-coated test piece surface.}
\label{fig:roughness_sample}
\end{figure}

For prototyping and characterization, $100\,\text{mm} \times 80\,\text{mm}$ aluminium test pieces were prepared. Each plate was first coated with a matte finish paint, then uniformly coated with adhesive, and subsequently sprinkled with a single, non-overlapping layer of glass beads for adherence. Due to the susceptibility of the beads to detachment, careful handling of these samples was imperative. Figure \ref{fig:roughness_sample} illustrates the applied glass beads, hereafter referred to as glass-DMR. Panel (a) provides a photograph of the beads adhered to an aluminium sample, whilst the upper figure of panel (b) presents a magnified image captured using a three-dimensional measuring laser microscope (Olympus LEXT OLS), and the lower figure of panel (b) displays the Probability Density Function (PDF) of the roughness element height. The measured roughness parameters for this glass-DMR on a flat aluminium plate were: arithmetic mean roughness ($R_a$, JIS 1994) = $3.678\,\mu\text{m}$, maximum height roughness ($R_y$) = $17.944\,\mu\text{m}$, and ten-point mean roughness ($R_z$, JIS 1994) = $17.5\,\mu\text{m}$.

\begin{figure}[H]
\centering
\subfigure[Magnified image of the cylinder surface]{
\includegraphics[width=0.45\textwidth]{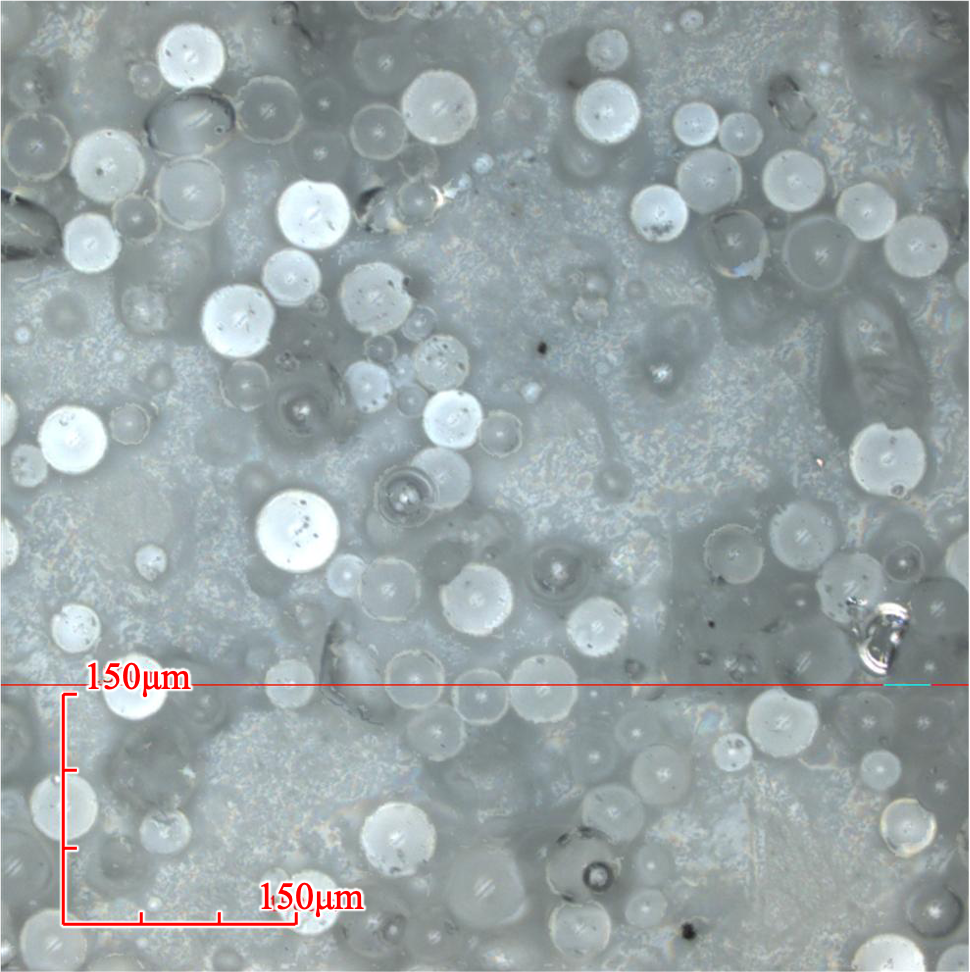}}
\hspace{2em}
\subfigure[Surface height characteristics; one-dimensional height profile measured along a horizontal line indicated by the red line in the corresponding surface image (upper) and Probability Density Function (PDF) of the roughness height, calculated from the one-dimensional height data shown in Panel (a) (below)]{
\includegraphics[width=0.46\textwidth]{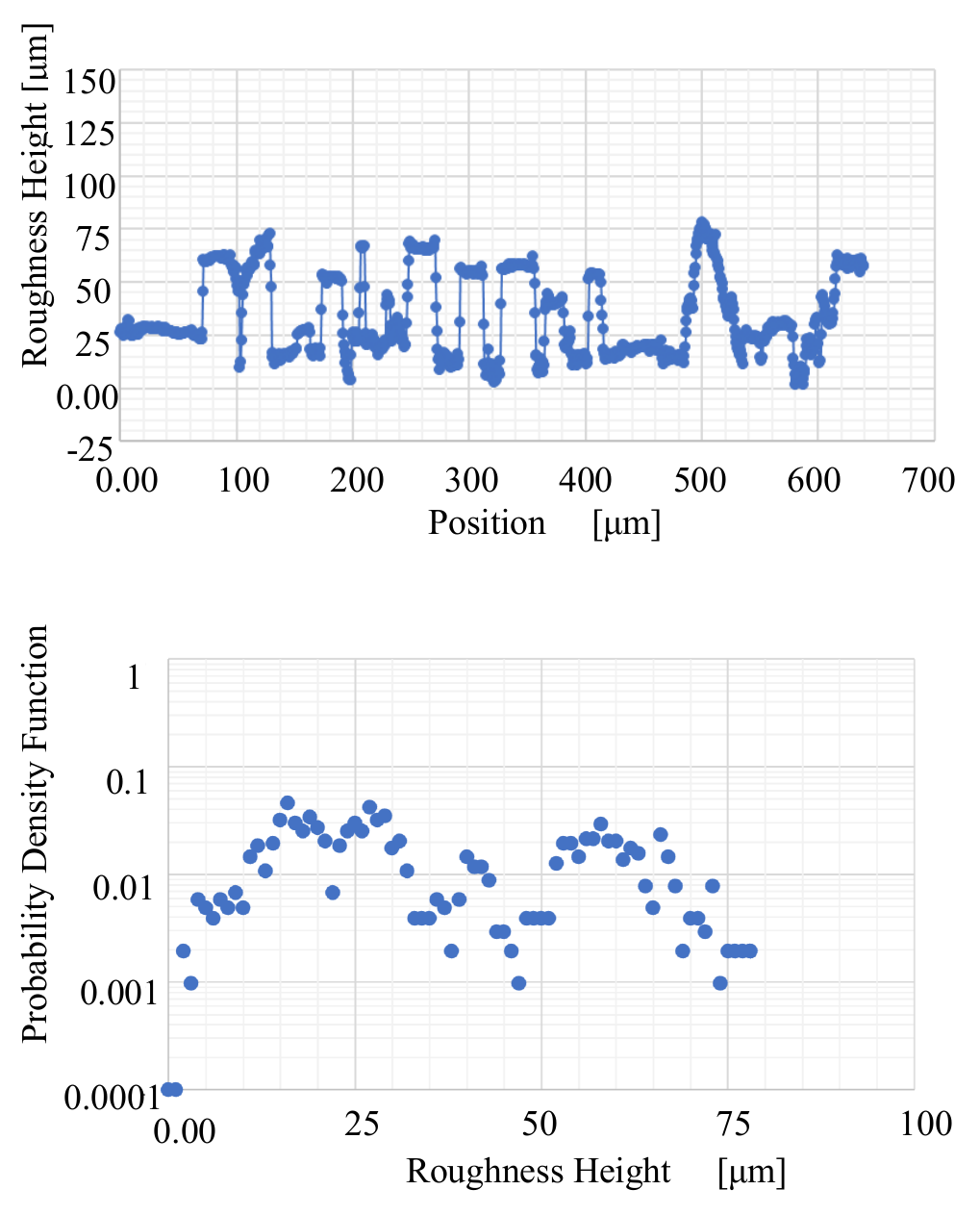}}
\caption{Characteristics of the glass-DMR-coated cylinder surface.}
\label{fig:roughness_cylinder}
\end{figure}

Figure \ref{fig:roughness_cylinder} presents the surface characteristics of the glass-DMR applied to the cylindrical streamlined model used in the main experiments. Initial measurements of the glass-DMR on the cylinder, obtained using a one-dimensional contact-type roughness profiler, yielded: an arithmetic mean roughness ($R_a$) of $11.81\,\mu\text{m}$, an average maximum height of $72.97\,\mu\text{m}$, and a Root Mean Square (RMS) roughness of $14.82\,\mu\text{m}$. Post-experiment measurements revealed a reduction in these parameters, with average values measuring: $R_a = 6.48\,\mu\text{m}$, average maximum height $= 48.6\,\mu\text{m}$, and average RMS roughness $= 8.68\,\mu\text{m}$, indicating cumulative bead loss during testing. Specifically, Panel (a) of Figure \ref{fig:roughness_cylinder} shows a magnified photograph of the surface post-experiment, acquired with the Olympus LEXT OLS, whilst Panel (b) displays the one-dimensional height profile along the red horizontal line indicated in Panel (a). The streamlined model, with its DMR coating, was reinstalled for each measurement run, which inevitably led to a minor, cumulative loss of beads with every setup. The manufacturing process ensured a random distribution of the roughness elements. While the element height is primarily governed by the bead diameter, the measured Probability Density Function (PDF) of the roughness element height is not uniform; instead, it exhibits a complex distribution with two distinct concentration peaks around $20\,\mu\text{m}$ and $60\,\mu\text{m}$, as illustrated in the lower figure of Figure \ref{fig:roughness_sample}(b).

\subsubsection{Phase II}

For the second phase of the experiment, the DMR coating was outsourced to O-Well Inc., a company with extensive experience in aerodynamic surface applications, notably their collaboration with Japan Airlines (JAL) and the Japan Aerospace Exploration Agency (JAXA) on Boeing 787 riblet coatings using their proprietary "paint-to-paint method" \citep{JAL_JAXA_Owell_Nikon_2023, JAL_JAXA_Owell_2023, JP6511612B2}. This commissioned DMR coating was specifically designed to replicate the unique roughness characteristics of the glass bead-based DMR described previously. It is important to clarify that this particular DMR application process for our test model differs slightly from the method used for the large-scale riblet demonstration on the B787. In our case, a specialised base coating was applied to the model, followed by a sandblasting process to create the desired concave DMR pattern.

The roughness parameters measured after coating were as follows: for DMR1, the arithmetic mean roughness ($Ra$) was $2.63\,\mu\text{m}$. The distribution of its maximum height roughness ($Ry$) was characterized by a mean of $36.6\,\mu\text{m}$, with a standard deviation of $4.5\,\mu\text{m}$ and a variance of $20.3\,\mu\text{m}^2$. For DMR2, the arithmetic mean roughness ($Ra$) was $2.82\,\mu\text{m}$. The distribution of its maximum height roughness ($Ry$) exhibited a mean of $52.0\,\mu\text{m}$, with a standard deviation of $5.2\,\mu\text{m}$ and a variance of $26.6\,\mu\text{m}^2$.

\begin{figure}[H]
\centering
\subfigure[DMR1 surface]{
\includegraphics[width=0.45\textwidth]{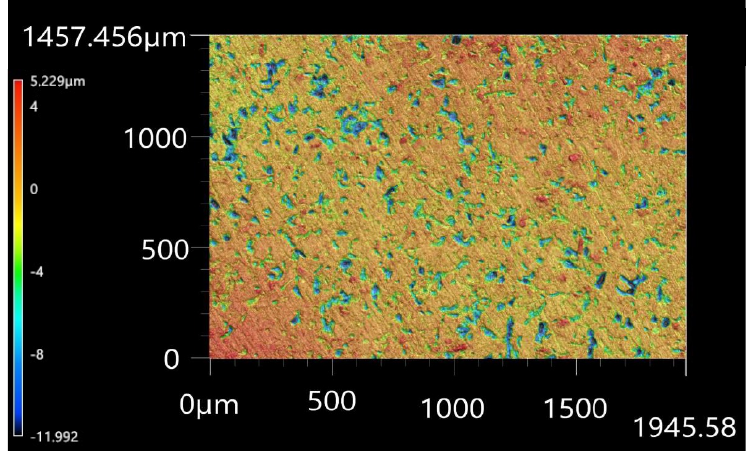}}
\hspace{2em}
\subfigure[DMR2 surface]{
\includegraphics[width=0.45\textwidth]{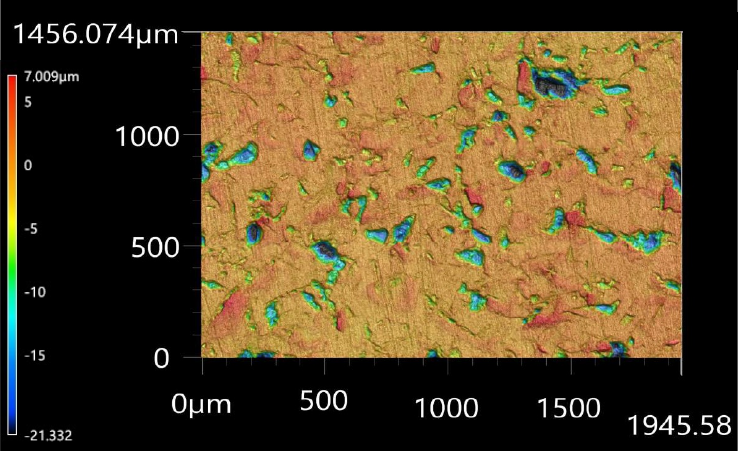}}
\caption{Characteristics of the DMR1 and DMR2 surfaces of a cylinder.}
\label{fig:roughness_dmr}
\end{figure}

\begin{figure}[H]
\centering
\subfigure[DMR1 surface]{
\includegraphics[width=0.45\textwidth]{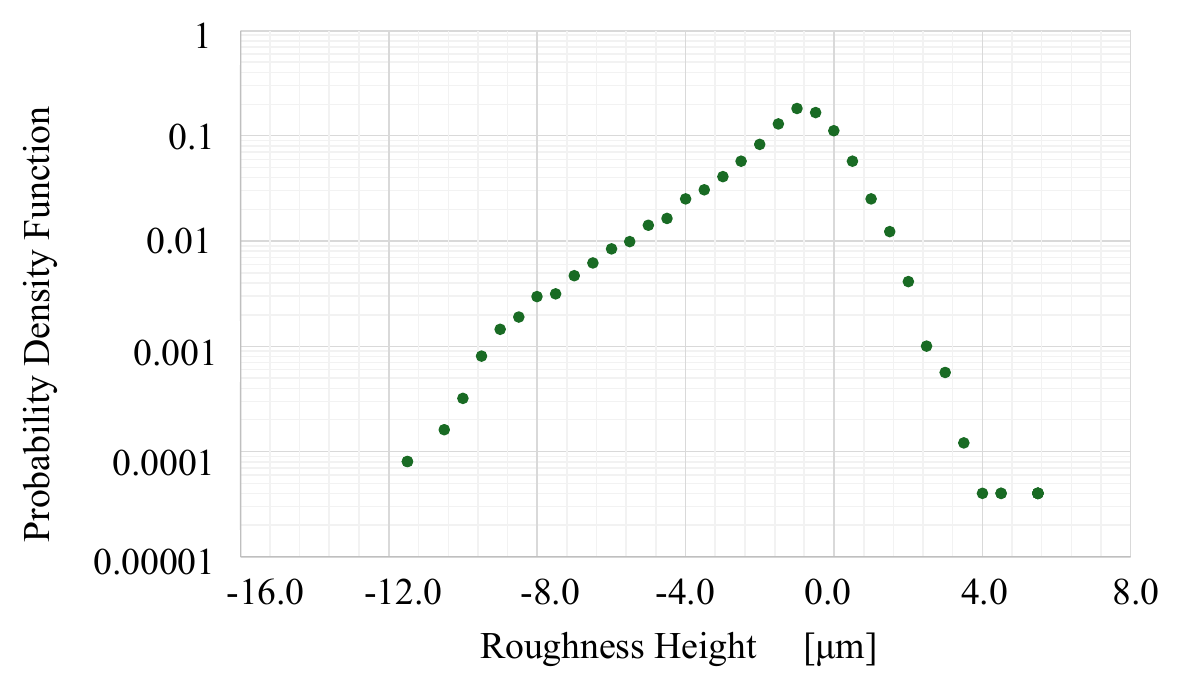}}
\hspace{2em}
\subfigure[DMR2 surface]{
\includegraphics[width=0.45\textwidth]{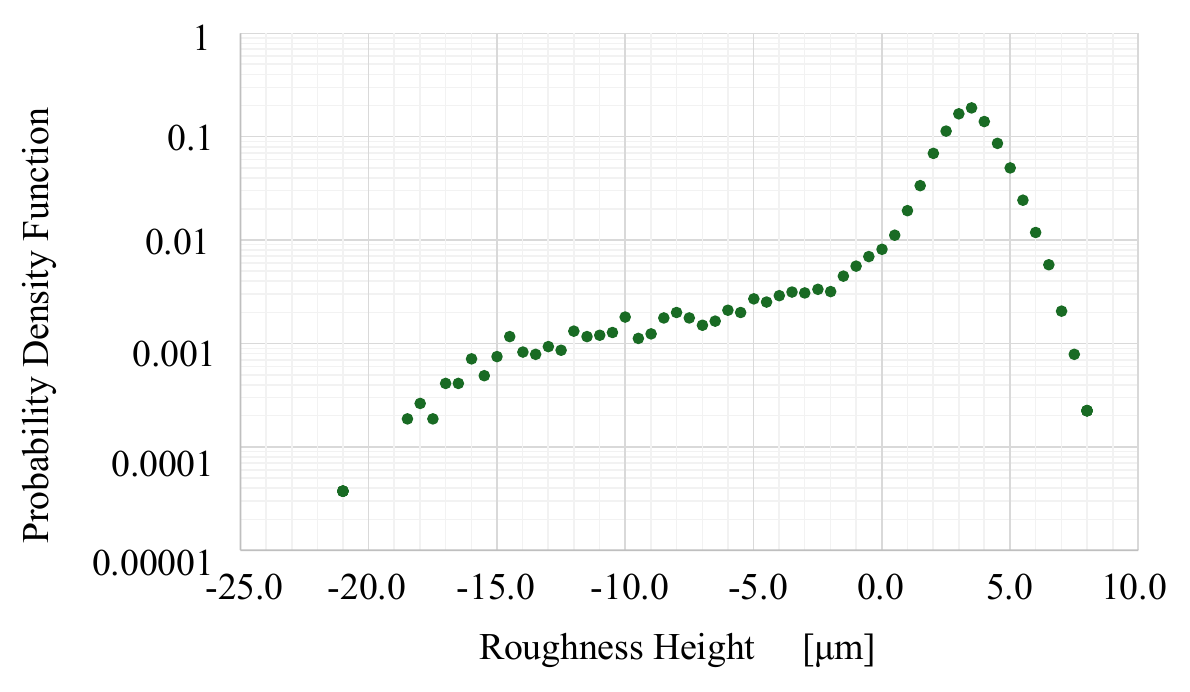}}
\caption{Probability Density Function (PDF) of the roughness height, calculated from the images in Figure \ref{fig:roughness_dmr}, for the DMR1 and DMR2 cylinder surfaces.}
\label{fig:roughness_dmr_pdf}
\end{figure}

The two-dimensional distribution of roughness height for DMR1 and DMR2 is presented in Figures \ref{fig:roughness_dmr}(a) and (b), respectively. The Probability Density Function (PDF) calculated from these figures is also shown in Figures \ref{fig:roughness_dmr_pdf}(a) and (b). A key distinction between the DMR1 and DMR2 surfaces is that DMR2 features slightly fewer and deeper depressions. In our previous DNS parametric study on flat plates, we found that the wavelength of roughness elements is most effective in influencing the flow field when it is approximately equivalent to the boundary layer thickness \citep{ogawanumerical, ogawa2024parametric}. However, given that the distance between depressions for both DMR1 and DMR2 was considerably smaller than the boundary layer thickness in this experiment, significant differences in their effects based solely on this wavelength parameter were not strongly anticipated.

\subsection{Computational Methodology: Large Eddy Simulation (LES)}

The Large Eddy Simulation (LES) results were incorporated into this study to provide critical validation for our experimental data and to aid in the interpretation of the results obtained from the Magnetic Suspension and Balance System (MSBS). Specifically, the LES was utilised for three main purposes: (i) to validate the experimental total drag coefficient ($C_D$) in the laminar regime, (ii) to estimate the contribution of skin friction drag ($C_f$) to the total drag ($C_D$) for our streamlined test body—a necessity for discussing the DMR effect in terms of friction reduction—and (iii) to estimate the local boundary layer thickness ($\delta$), a critical flow parameter unobtainable via hot-wire or PIV measurements due to the constraints of the MSBS facility.

Simulations were performed using the open-source CFD toolbox, OpenFOAM (version v2406). The \texttt{pimpleFoam} solver, a transient, incompressible flow solver, was employed. The incompressible mass and momentum conservation equations were solved. To maintain computational efficiency, the solver was configured with one outer iteration per time step, which was made possible by setting a sufficiently small time step ($\Delta t = 1 \times 10^{-6}\,\text{s}$) corresponding to a maximum CFL number of $0.21$. The duration required to reach the initial time step was $0.36\,\text{s}$, and the average duration per time step in the main production runs was $0.22\,\text{s}$. For turbulence closure, Large Eddy Simulation (LES) with the Wall-Adapting Local Eddy-viscosity (WALE) model was utilised. For spatial discretisation, a linear upwind scheme was adopted for the convection term of the momentum equation to suppress numerical oscillations on the unstructured mesh, while a linear scheme (second-order central differencing) was applied to the remaining terms.

\begin{figure}[H]
\centering
\vspace{2em}
\subfigure[Overall view]{
\includegraphics[width=0.85\textwidth]{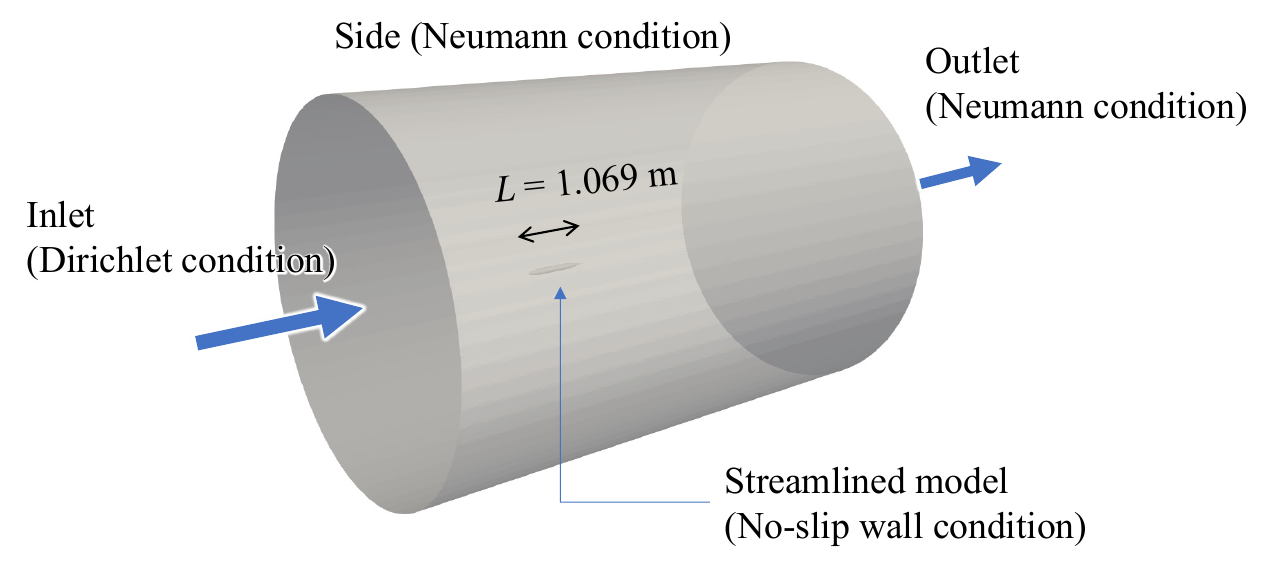}}
\subfigure[Cross-sectional view]{
\includegraphics[width=0.5\textwidth]{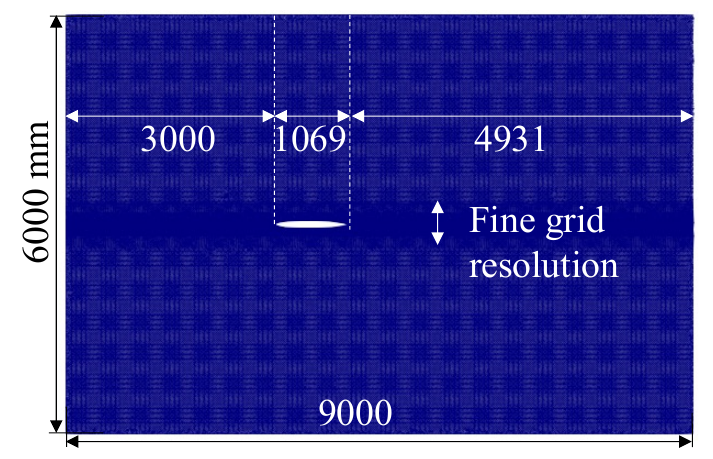}}
\caption{Computational domain and boundary conditions.}
\label{fig:les_domain}
\end{figure}

Simulations were conducted for three free-stream velocities, $U_{\infty}$: $10$, $25$, and $50\,\text{m/s}$. Corresponding free-stream conditions included a reference pressure $p_{\infty} = 0\,\text{kPa}$ and a temperature $T_{\infty} = 300\,\text{K}$. These conditions resulted in Reynolds numbers ($Re = U_{\infty}L/\nu$, where $L$ is the model total length, $1.069\,\text{m}$) of $7.1 \times 10^5$, $1.8 \times 10^6$, and $3.6 \times 10^6$, respectively. The computational domain and boundary conditions are depicted in Figure \ref{fig:les_domain}(a) and (b). The cylindrical computational domain measured $6.0\,\text{m}$ in diameter. This large domain size, resulting in a model blockage ratio of approximately $0.028\%$ based on the model's maximum frontal area, was deliberately chosen to minimize the unphysical influence of the outer boundary conditions and to mitigate the artificial flow constriction (blockage effect) around the model, ensuring the simulation closely approximates free-stream conditions. The upstream distance from the inlet boundary to the model's leading edge was $3.0\,\text{m}$, whilst the downstream distance from the model's trailing edge to the outlet boundary was $4.9\,\text{m}$. Due to limitations in computational resolution and cost, the tripping tapes used in the experimental setup were not explicitly resolved in the simulations.

\begin{figure}[H]
\centering
\includegraphics[width=0.5\textwidth]{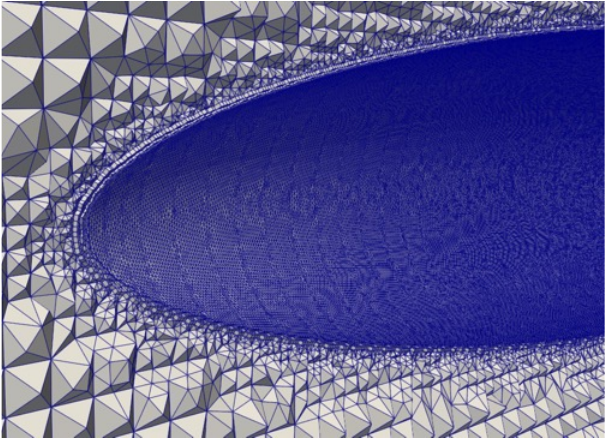}
\caption{Enlarged view of computational grid around the leading edge of the baseline simulations.}
\label{fig:les_grid2}
\end{figure}

The specific boundary conditions are detailed as follows: A Dirichlet condition was applied for velocity at the inlet boundary. A Neumann condition (zero-gradient for velocity) was applied to the side wall of the cylindrical computational domain, simulating a slip wall. At the outlet boundary, a zero-gradient Neumann condition was imposed. Finally, a no-slip wall condition was imposed on the streamlined model surface. To accurately resolve the flow features around the streamlined model and within its wake, regions of high mesh density were implemented, as depicted in Figure \ref{fig:les_domain}(b). A boundary layer mesh, consisting of 5 layers in the wall-normal direction with a first layer height of $0.00005\,\text{m}$, was incorporated near the model surface. The wall-surface-averaged values for the first near-wall resolution in wall units, $y_1^+$, were $0.56$, $1.0$, and $1.8$ for the cases of $10\,\text{m/s}$, $25\,\text{m/s}$, and $50\,\text{m/s}$, respectively, ensuring near-wall turbulence is adequately resolved. Figure \ref{fig:les_grid2} further illustrates an enlarged view of the computational mesh around the leading edge. The total cell count for the mesh was $21,231,139$.

In addition to the baseline simulations consisting of 21,231,139 cells and five boundary layers (with a first-layer height of $0.00005\,\text{m}$), an additional set of high-resolution simulations was performed to further ensure the fidelity of the boundary layer state and pressure recovery at the tail. For these refined cases, the total cell count was increased to 45,384,172, and the boundary layer mesh was expanded to 10 layers with a reduced first-layer height of $0.00002\,\text{m}$. The resulting wall-surface-averaged $y_1^+$ values were $0.20$, $0.39$, and $0.65$ for the cases of $10\,\text{m/s}$, $25\,\text{m/s}$, and $50\,\text{m/s}$, respectively, indicating that the near-wall region was fully resolved. For the $50\,\text{m/s}$ case, the time step was set to $\Delta t = 1 \times 10^{-7}\,\text{s}$, corresponding to a maximum CFL number of $0.55$. The total physical simulation time reached $1.0\,\text{s}$, with the time-averaging for drag statistics performed over the final $0.44\,\text{s}$ of the run. Furthermore, the \texttt{nutkWallFunction} was applied to the turbulent kinematic viscosity ($\nu_t$) at the model surface. This wall function provides a unified profile by blending analytical expressions for the viscous sublayer and the logarithmic region based on the local turbulent kinetic energy ($k$). Even with our fine mesh resolution ($y_1^+ < 0.65$), this approach ensures a robust and consistent prediction of the wall shear stress by accounting for the local boundary layer state, thereby enhancing the reliability of the skin friction drag ($C_f$) estimation across the investigated Reynolds number range.

\section{Validation of Measurements\label{sec:measurement}}

\subsection{Measurement Error}
Accurate and precise measurements of forces and moments are paramount for the validation of our experimental results. Measurement error was quantified by assuming a linear relationship between the applied external load and the corresponding output current from the MSBS. A regression line was established using the least squares method, based on the gradient derived from the output current. The error was subsequently defined as the maximum deviation between the external force calculated from this output current and the actual applied load on the model.

Measurement resolution was determined by considering both the amplifier's gain ($\text{A/V}$) and the resolution of the digital-to-analogue ($\text{D/A}$) converter ($\text{V/count}$). Through force calibration tests, the magnetic force generated per unit current ($\text{N/A}$ or $\text{Nm/A}$) was obtained. This value was then utilised to calculate the magnetic force generated per $\text{D/A}$ converter count ($\text{N/count}$ or $\text{Nm/count}$), thereby characterizing the system's resolution. The measurement errors in drag, alongside the full-scale ($\text{F.S.}$) values and resolutions for various force and moment components, are presented in Table \ref{tab:error}. As detailed in the table, for the streamlined model (Phase II) employed in this study, the drag measurement resolution was $0.000899\,\text{N/count}$, with an associated error of $0.36\%\,\text{F.S.}$ (where $\text{F.S.}$ corresponds to $2.87\,\text{N}$). It should be noted that the Full Scale (F.S.) values presented in Table \ref{tab:error} represent the maximum static load applied to the model during the pre-test force calibration. The F.S. value of $2.87\,\text{N}$ for the streamlined model was intentionally set to be approximately twice the maximum expected drag force (recorded as $1.3\,\text{N}$ for the Phase I glass-DMR surface with tripping tapes at $50\,\text{m/s}$). This proactive setting was chosen to account for potential transient or fluctuating aerodynamic forces (dynamic loading) that could occur during the experiment.

\begin{table}
\begin{center}
\def~{\hphantom{0}}
\begin{tabular}{llll}
\multicolumn{1}{c}{\textbf{Case}} & \multicolumn{1}{c}{\textbf{Error [\% F.S.]}} & \multicolumn{1}{c}{\textbf{F.S. [N or Nm]}} & \multicolumn{1}{c}{\textbf{Resolution [N/count or Nm/count]}} \tabularnewline
\addlinespace[1ex]
Drag (Cylinder) & $0.18\,\%$ & $4.22\,\text{N}$ & $0.001527\,\text{N/count}$ \tabularnewline
Drag (AGARD-B model) & $0.17\,\%$ & $6.18\,\text{N}$ & $0.002516\,\text{N/count}$ \tabularnewline
Drag (Present streamlined model) & $0.36\,\%$ & $2.87\,\text{N}$ & $0.000899\,\text{N/count}$ \tabularnewline
Side force (AGARD-B model) & $0.25\,\%$ & $3.24\,\text{N}$ & $0.004907\,\text{N/count}$ \tabularnewline
Lift (AGARD-B model) & $0.26\,\%$ & $29.49\,\text{N}$ & $0.009697\,\text{N/count}$ \tabularnewline
Pitching moment (AGARD-B model) & $0.53\,\%$ & $0.17\,\text{Nm}$ & $0.003717\,\text{Nm/count}$ \tabularnewline
Yawing moment (AGARD-B model) & $0.62\,\%$ & $0.76\,\text{Nm}$ & $0.001783\,\text{Nm/count}$ \tabularnewline
Rolling moment (AGARD-B model) & $0.12\,\%$ & $0.02\,\text{Nm}$ & $0.000018\,\text{Nm/count}$ \tabularnewline
\end{tabular}
\caption{\label{tab:error}Error estimation and resolution of 1-m MSBS measurement system.}
\end{center}
\end{table}

\subsection{Stability of the Model Position and Attitude}
The stability of the model's position and attitude was meticulously monitored throughout the experiments. The key operational and control limits of the 1-m MSBS system are provided in Table \ref{tab:msbs_parameters}. These parameters govern the model's allowed movement before control action is taken or the system shuts down. It is important to note that the values listed for Allowed Max Displacement ($\pm 0.06\,\text{mm}$) and Allowed Max Velocity ($\pm 0.005\,\text{mm/s}$) correspond precisely to the maximum observed amplitude and velocity of model oscillation, respectively, when the model is held stationary during wind-off and wind-on conditions. These maximum values are taken from the entirety of the position and attitude data. Crucially, the amplitude and velocity values presented in Table \ref{tab:msbs_parameters} (maximum single point values) were derived from the position and attitude data after the application of a low-pass filter.

Table \ref{tab:position_attitude_stability}, on the other hand, summarizes the standard deviations and mean values of the model's position and attitude variation components for the present measurement conditions. These standard deviations are calculated from position and attitude data (8191 points) measured over $6.55$ seconds at a sensing frequency of $1250\,\text{Hz}$. The standard deviations presented in Table \ref{tab:position_attitude_stability} were calculated using the raw position and attitude data, which includes high-frequency noise. If the standard deviation for the data in Table \ref{tab:position_attitude_stability} were calculated after applying the low-pass filter, the magnitude of the low-pass filtered standard deviations were found to be $x: 0.01\,\text{mm}$, $y: 0.015\,\text{mm}$, $z: 0.015\,\text{mm}$ for translational components, and $t: 0.005\,\text{deg.}$, $p: 0.005\,\text{deg.}$ for rotational components. These results unequivocally confirm that the model maintained exemplary stability during the tests, with vibrations remaining well within limits for highly accurate aerodynamic force measurements.

\begin{table}
\centering
\begin{tabular}{lcl}
\textbf{Parameter} & \textbf{Value} & \multicolumn{1}{c}{\textbf{Description}} \tabularnewline
\addlinespace[1ex]
Allowed Max Displacement & $\pm 0.06\,\text{mm}$ & Maximum position deviation allowed by the control system \tabularnewline
& & in a stationary state. \tabularnewline
\addlinespace[0.5ex]
Allowed Max Velocity & $\pm 0.005\,\text{mm/s}$ & Maximum velocity allowed by the control system, \tabularnewline
& & corresponding to the maximum observed vibration velocity \tabularnewline
& & in a stationary state. \tabularnewline
\addlinespace[0.5ex]
Sensing Frequency & $1250\,\text{Hz}$ & Readout frequency of the line sensor cameras. \tabularnewline
\end{tabular}
\caption{Key operational and control limits of the 1-m Magnetic Suspension and Balance System (MSBS), when the model is held stationary during wind-off and wind-on conditions.}
\label{tab:msbs_parameters}
\end{table}

\begin{table}
\begin{center}
\def~{\hphantom{0}}
\begin{tabular}{llll}
& \multicolumn{1}{c}{\textbf{Component}} & \multicolumn{1}{c}{\textbf{Standard Deviation}} & \multicolumn{1}{c}{\textbf{Mean of Variation Component}} \tabularnewline
\addlinespace[1ex]
& $x$ & $0.02\,\text{mm}$ & $3.5\times 10^{-18}\,\text{mm}$ \tabularnewline
& $y$ & $0.04\,\text{mm}$ & $1.1\times 10^{-19}\,\text{mm}$ \tabularnewline
& $z$ & $0.04\,\text{mm}$ & $5.4\times 10^{-19}\,\text{mm}$ \tabularnewline
& $t$ & $0.01\,\text{deg}$ & $3.9\times 10^{-16}\,\text{deg}$ \tabularnewline
& $p$ & $0.02\,\text{deg}$ & $5.7\times 10^{-16}\,\text{deg}$ \tabularnewline
\end{tabular}
\caption{\label{tab:position_attitude_stability}Standard deviation and mean value of position/attitude variation, for the experimental condition of $U = 11.2\,\text{m/s}$ and $Re = 807,800$ (based on raw data).}
\end{center}
\end{table}

As detailed in the table, for the experimental condition of $U = 11.2\,\text{m/s}$ and $Re = 807,800$, the mean values of the variation components for both translational ($x$, $y$, $z$) and rotational (yaw, pitch, and roll, denoted as $t$, $p$, and $r$ respectively in MSBS systems) degrees of freedom were exceedingly close to zero. This signifies a negligible average displacement from the nominal position and attitude. The standard deviations, which quantify the magnitude of fluctuations, were found to be small, with maximum values of $0.04\,\text{mm}$ for translational motion (in the $y$ and $z$-directions) and $0.02\,\text{degrees}$ for rotational motion (about the $p$-axis). This is a crucial aspect when considering the potential joint effects of model vibration and surface roughness.

\begin{figure}[H]
\centering
\subfigure[Plain case]{
\includegraphics[width=0.98\textwidth]{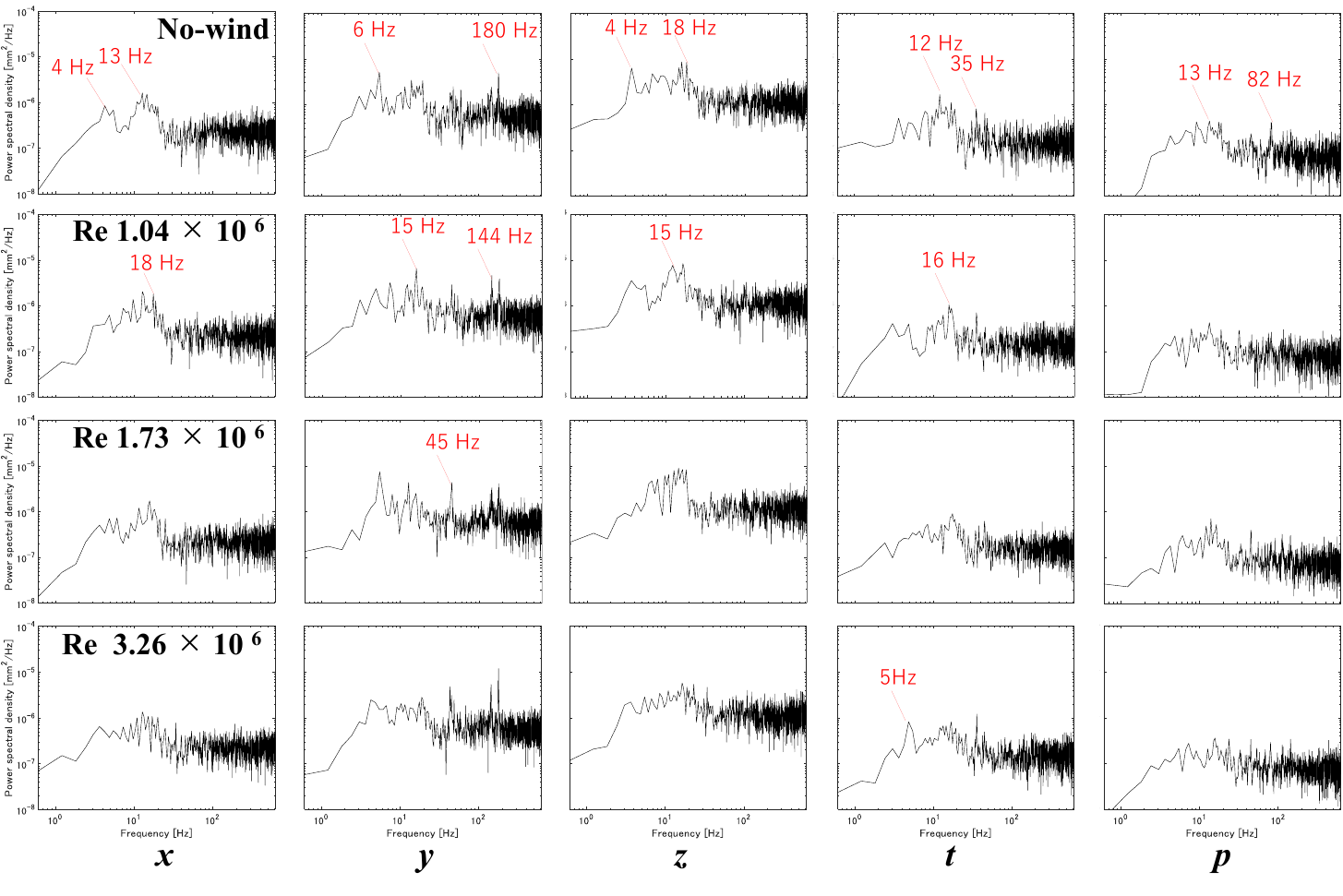}}
\subfigure[glass-DMR case]{
\includegraphics[width=0.98\textwidth]{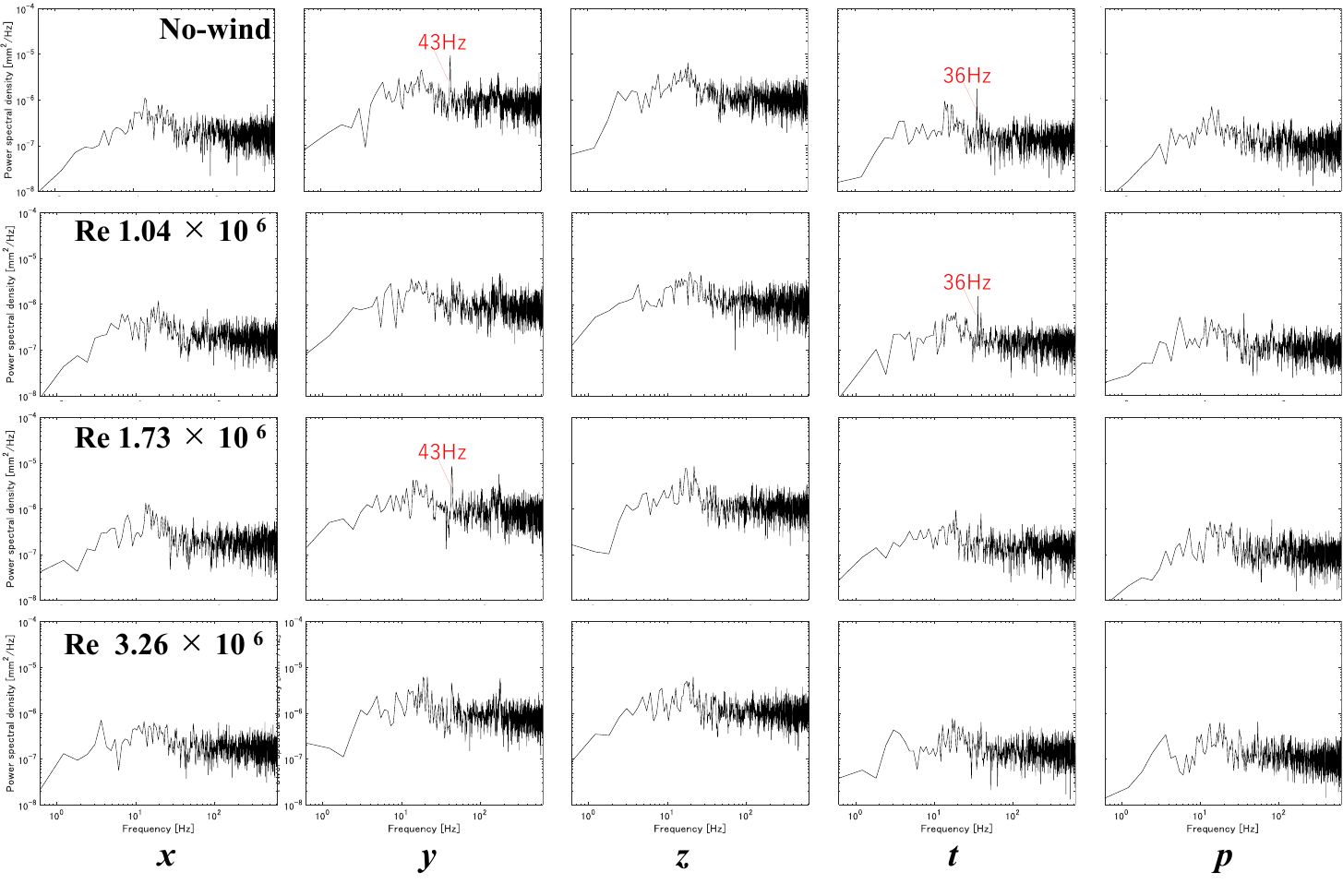}}
\caption{Power Spectral Density (PSD) of the model's position and attitude variation components measured by the MSBS position sensor. Figures show the results (a) for the Plain case and (b) for the glass-DMR case. The data presented is the raw, unfiltered signal, illustrating the full spectrum of vibration noise, including high-frequency components from the sensing system. The PSD provides detailed insight into the frequency distribution of the model's movements across the five degrees of freedom.}
\label{fig:frequency}
\end{figure}

The Power Spectral Densities (PSDs) of the position and attitude data for the Plain and glass-DMR cases (Phase I, unfiltered) are shown in Figure \ref{fig:frequency}(a) and (b), respectively, confirming the frequency distribution of the observed vibrations. The prominent peaks are generally concentrated in the low-frequency range, and notably, all frequencies presented in this figure are substantially lower than those relevant to boundary layer transition. Based on boundary layer thicknesses computed by LES, the characteristic frequencies associated with Tollmien-Schlichting (TS) transition are estimated to be in the order of $200\,\text{Hz}$ to $3000\,\text{Hz}$ \citep{mack1984boundary, schlichting2017boundary}, which is significantly higher than the main peaks observed in the MSBS position spectrum. It is acknowledged that the maximum raw standard deviation of $0.04\,\text{mm}$ ($40\,\mu\text{m}$) for translational displacement is of the same order of magnitude as the roughness height ($R_y \approx 48.6\,\mu\text{m}$). Therefore, the influence of model vibrations on flow transition over the DMR surfaces cannot be entirely ruled out. However, it was confirmed that the frequency spectra distribution remained almost identical across all tested flow velocities and levitation cases. This consistency indicates that the conditions pertaining to any potential vibrational influence on transition were comparable throughout the experimental campaign.

%============================================================================
% Section 3.3 Comparison of LES Results...
%============================================================================

\subsection{Comparison of LES Results with Experimental Results without Tripping (Phase I)\label{sec:les}}

Aerodynamic drag force was initially measured for both the smooth and glass-DMR coated streamlined models without the application of tripping tapes. A Cosmo Instruments model DM-3501 differential pressure gauge ($5\,\text{kPa}$ range) was employed for these measurements. Data acquisition involved sweeping the flow velocity from low to high and then back, constituting one `run'. Drag values were recorded over two such runs (Run 1 and Run 2), yielding four data points per condition (an up-sweep and a down-sweep for each run). Each data acquisition represented a time average over $6.553\,\text{s}$ ($8191$ points at $1250\,\text{Hz}$).

\begin{figure}[H]
\centering
\includegraphics[width=1.0\textwidth]{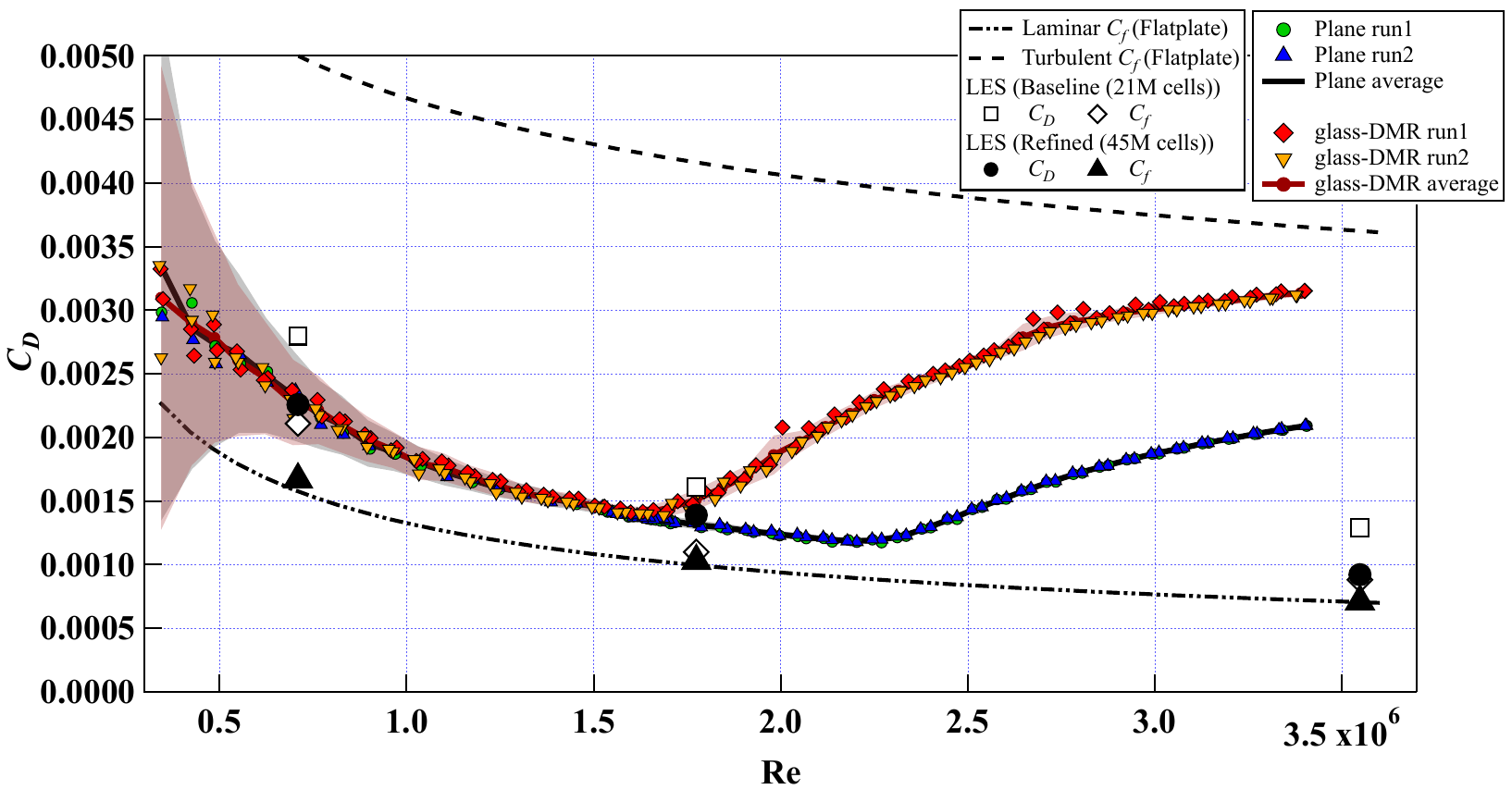}
\caption{Comparison of the total drag coefficient ($C_D$) versus the Reynolds number ($\mathit{Re}$) for the smooth and glass-DMR surfaces in Phase I (without tripping tapes). The data include results from Baseline (21M cells) and Refined (45M cells) Large Eddy Simulation (LES) and experiments using the Magnetic Suspension and Balance System (MSBS). For the LES Results, open square ($\square$) and diamond ($\diamond$) symbols denote baseline $C_D$ and $C_f$, respectively, while solid circle ($\bullet$) and triangle ($\blacktriangle$) symbols denote refined $C_D$ and $C_f$, respectively. The dashed lines represent the theoretical skin friction coefficient ($C_f$) for laminar and turbulent flow. For the MSBS Experimental Data of the Smooth Surface, small circular ($\circ$) and triangular ($\triangle$) symbols show individual runs (Run 1 and Run 2), and the thick solid line represents the averaged total drag. For the glass-DMR Surface, small diamond ($\diamond$) and inverted-triangular ($\nabla$) symbols show individual runs (Run 1 and Run 2), and circular symbols with a thick solid line represent the averaged total drag for this surface.}
\label{fig:no_trip_comparison}
\end{figure}

The results of the drag decomposition and experimental comparison for the smooth streamlined model are summarised in table~\ref{tab:drag_decomposition_summary}. The experimental results are further presented in figure~\ref{fig:no_trip_comparison}, where `average' denotes the mean value of Run 1 and Run 2. The aerodynamic drag force, directly calculated from the electromagnetic force, was normalised by the nozzle differential pressure and the model's reference surface area ($S = 0.26885663\,\text{m}^2$). The resultant drag coefficient ($C_D$) is shown on the vertical axis. The $95\,\%$ coverage uncertainty, indicated by the shaded region, was calculated from the measurement accuracy and the random error derived from the four data points. Measurement accuracy considered three error sources: force calibration ($0.017880075\,\text{N}$ for a maximum load of $1.79\,\text{N}$ in the measurement plane); the differential pressure gauge ($8.5\,\text{Pa}$ for the DM-3501 $5\,\text{kPa}$ model); and surface area ($0.00000001\,\text{m}^2$, assuming a manufacturing tolerance of $0.10\,\text{mm}$). These contributions were computed using their respective sensitivities.

\begin{table}
  \begin{center}
\def~{\hphantom{0}}
  \begin{tabular}{lccccc}
      Method  & Surface & $\mathit{Re}$ & $C_D$ & $C_f$ & $C_p$ \\ [3pt]
      Theory (Laminar/Flat plate) & Smooth & $3.6 \times 10^6$ & - & 0.000725 & - \\
      LES (Laminar/Streamline)   & Smooth & $3.6 \times 10^6$ & 0.000922 & 0.000712 & 0.00021 \\
      Theory (Laminar/Flat plate) & Smooth & $1.8 \times 10^6$ & - & 0.00100~ & - \\
      LES (Laminar/Streamline)   & Smooth & $1.8 \times 10^6$ & 0.00139~ & 0.00103~ & 0.00036 \\
      MSBS (Experiment)           & Smooth & $1.8 \times 10^6$ & 0.00131~ & - & - \\
      Theory (Laminar/Flat plate) & Smooth & $0.71 \times 10^6$ & - & 0.00153~ & - \\
      LES (Laminar/Streamline)   & Smooth & $0.71 \times 10^6$ & 0.00226~ & 0.00167~ & 0.00059 \\
      MSBS (Experiment)           & Smooth & $0.71 \times 10^6$ & 0.00232~ & - & - \\
  \end{tabular}
  \caption{Comparison of drag coefficients between theoretical predictions for a smooth flat plate, numerical simulations (refined LES, Laminar/Streamline), and MSBS experiments for the smooth streamlined model at various Reynolds numbers.}
  \label{tab:drag_decomposition_summary}
  \end{center}
\end{table}

For cases without tripping tapes, the glass-DMR coating was observed to initiate a significant drag rise at a lower flow velocity compared to the smooth surface (Plain). This indicates that the glass-DMR effectively promotes boundary layer transition, a commonly known effect of distributed micro-roughness (DMR) frequently utilised in wind tunnel experiments to encourage transition and achieve fully turbulent flow more smoothly.

To provide a rigorous quantitative baseline, two sets of LES results are presented in figure \ref{fig:no_trip_comparison}: the baseline simulations (21M cells) and the refined simulations (45M cells). In the legend, LES (Baseline) values for the total drag coefficient ($C_D$) and skin friction coefficient ($C_f$) are indicated by open squares ($\square$) and open diamonds ($\diamond$), respectively. The LES (Refined) values, which resolve the near-wall region with $y_1^+ < 1.0$, are denoted by solid circles ($\bullet$) for $C_D$ and solid triangles ($\blacktriangle$) for $C_f$. The upper and lower dashed lines correspond to theoretical values for the skin friction drag coefficient for a flat plate under fully turbulent and laminar boundary layer conditions, respectively. The theoretical formulae employed are $C_f\,\text{(Laminar)} = 1.328 / \sqrt{Re}$ and $C_f\,\text{(Turbulent)} = 0.0592 / (Re)^{0.2} / 0.8$.

Notably, the $C_f$ values obtained from both LES cases are quite close to the theoretical values of $C_f$ for a laminar boundary layer on a flat plate. Particularly for the refined LES, the predicted $C_f$ at $Re = 3.6 \times 10^6$ matches the theoretical Blasius solution with an error of less than $1.0\%$. This level of agreement confirms that the simulation accurately resolves the laminar flow field without premature numerical transition, thereby providing a reliable "Laminar Limit" for comparison with experimental data where natural transition occurs.

It is essential to analyse the pressure drag coefficient ($C_p$), defined as the difference between $C_D$ and $C_f$. This component represents the integral of the normal stress acting over the model surface and is primarily governed by the inherent high pressure in the nose stagnation region and the pressure recovery achieved by the attached flow at the aft section. The grid refinement significantly improved the pressure recovery estimation at the aft section, leading to a further reduction in the estimated $C_p$ compared to the baseline results. The LES results indicate that $C_p$ remains small irrespective of the flow velocity, ranging from $0.00059$ to $0.00021$ in the refined cases. This value corresponds to $22.8\%$ to $26.1\%$ of the total drag but is consistently less than half the magnitude of $C_f$. This constancy of the pressure component is further supported by the MSBS experimental measurements, where the difference between the measured total drag ($C_D$) and the theoretical laminar skin friction ($C_f$), which equals to pressure drag ($C_p$), stays small.

Furthermore, experimental evidence reinforces that the DMR does not change drag by suppressing separation. In the low Reynolds number regime shown in figure \ref{fig:no_trip_comparison}, the measured $C_D$ values for the smooth and glass-DMR surfaces are nearly identical. If the DMR were reducing pressure drag by suppressing wake separation (similar to a dimpled sphere), a significant difference in $C_D$ would be expected in this regime where separation is most likely. The absence of such a deviation confirms that the macroscopic flow topology is consistent between the surfaces.

Most importantly, the refined LES quantitatively establishes that the total available pressure drag budget at $Re = 3.6 \times 10^6$ is only approximately $0.00021$. As discussed in the following sections, given that the total drag modification observed in the Phase II experiments (with tripping) is maximally on the order of $0.001$, even a theoretical $100\%$ elimination of pressure drag at the tail could only account for approximately $20\%$ of the observed benefit. This quantitative discrepancy serves as physical proof that the primary mechanism for the substantial drag modification must be due to the change of skin friction drag ($C_f$) through the modification of the boundary layer, rather than the suppression of separation ($C_p$).

%============================================================================
% 					        Result
%============================================================================
\section{Results and Consideration\label{sec:results}}
\subsection{Results for Cases with Tripping Tape 1 (Phase I)}
\begin{figure}[H]
\centering
\includegraphics[width=1.0\textwidth]{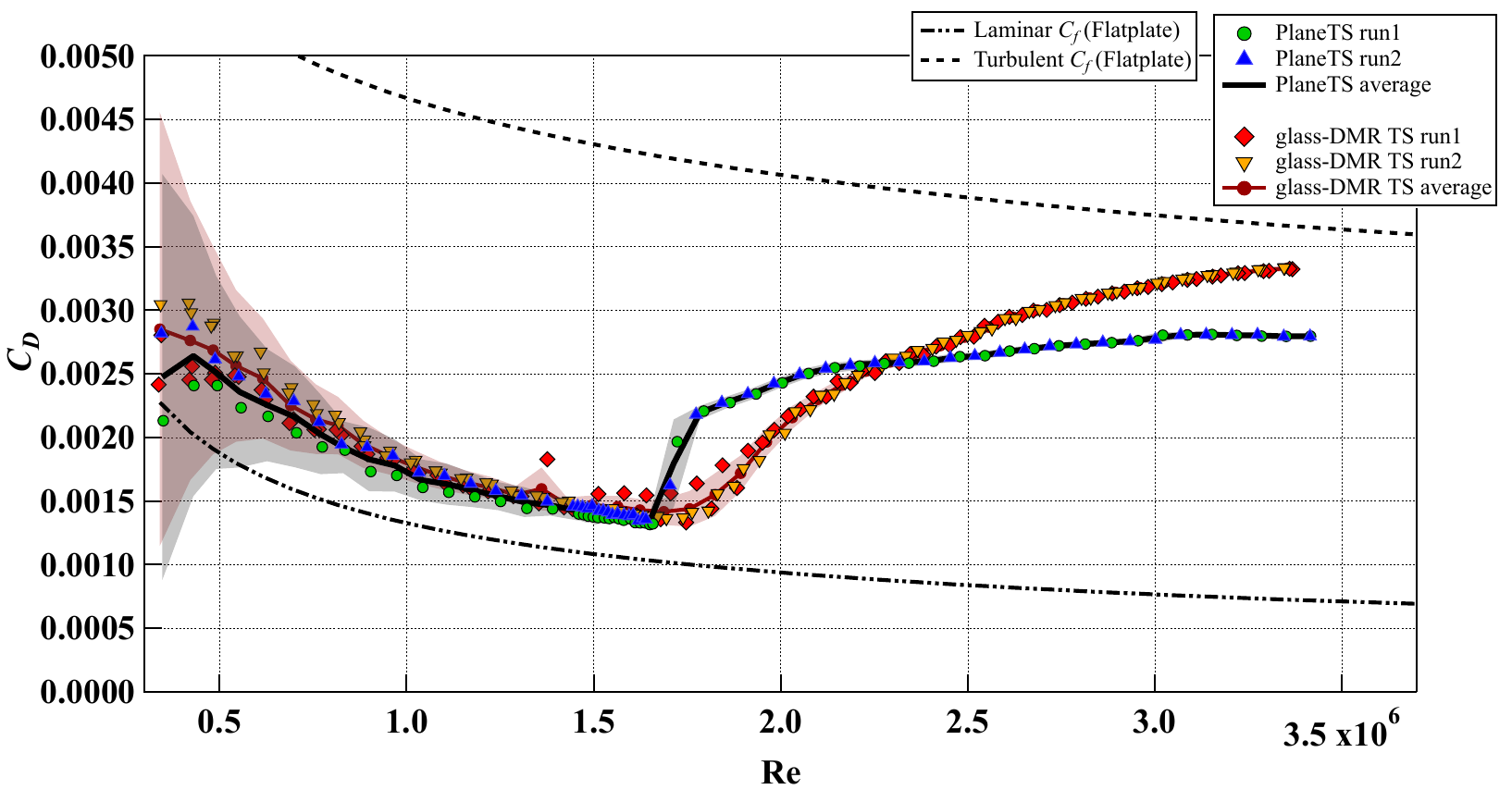}
\caption{Comparison of the total drag coefficient ($C_D$) versus the Reynolds number ($\mathit{Re}$) for the smooth (Plain) and glass-DMR surfaces in Phase I with tripping tapes applied. All data represent experimental results obtained using the Magnetic Suspension and Balance System (MSBS). The dashed lines represent the theoretical skin friction coefficient ($C_f$) for laminar and turbulent flow (e.g., flat plate formulae). For the MSBS Experimental Data of the Smooth Surface (Plain), small circular ($\circ$) and triangular ($\triangle$) symbols show individual runs (Run 1 and Run 2), and the thick solid line represents the averaged total drag. For the glass-DMR Surface, small diamond ($\diamond$) and inverted-triangular ($\nabla$) symbols show individual runs (Run 1 and Run 2), and circular symbols with a solid line represent the averaged total drag for this surface.}
\label{fig:Plane0}
\end{figure}

The aerodynamic drag force was measured with Tripping Tape 1 applied, as depicted in Figure \ref{fig:tape}(a). The experimental procedure and data processing methodologies remained identical to those detailed in the preceding section. Results are shown in Figure \ref{fig:Plane0}. A notable observation was a shift in the transition Reynolds number; however, the drag coefficient ($C_D$) values in both the low-speed laminar region and the high-speed turbulent region remained unaltered when compared to the corresponding cases without the tape. This consistently confirms that the presence or absence of the tripping tapes does not modify the $C_D$ values in these fully laminar or turbulent flow regimes.

In scaled wind tunnel testing, particularly for models such as aircraft wings, the upstream placement of tripping tape or discrete roughness elements is a conventional practice employed to ensure a fully turbulent flow state. This approach is widely considered appropriate for reproducing flight-scale flow fields in smaller wind tunnels where model dimensions are constrained. In the present study, the artificial disturbance of Tripping Tape 1 was applied and subsequently complemented by a glass DMR coating on a streamlined model, the onset Reynolds number for drag rise, referred to as the critical Reynolds number, increased compared to that observed on a smooth surface under identical conditions. Consequently, a substantial drag reduction of up to $34.4\%$ was achieved at approximately $Re = 1.78 \times 10^6$ within the transition region. Given that the flow field incorporating artificial disturbances in wind tunnel experiments more accurately emulates real-world conditions, these findings strongly suggest that the DMR coating holds significant promise for inducing a beneficial drag reduction effect in actual flow environments.

\subsection{Results for Cases with Tripping Tape 2 (Phase II)\label{sec:MSBS_PhaseII}}

In the second phase of experiments, two differential pressure gauges of Cosmo Instruments model DM-3501, with $500\,\text{Pa}$ and $5\,\text{kPa}$ ranges respectively, were employed to ensure highly precise wind speed measurements across the entire velocity spectrum. The $500\,\text{Pa}$ range gauge covered Reynolds numbers from $0.35 \times 10^6$ to $1.8 \times 10^6$, whilst the $5\,\text{kPa}$ range gauge was utilised for Reynolds numbers from $1.6 \times 10^6$ to $3.6 \times 10^6$. To ensure data consistency and provide an overlap between the two gauge ranges, Reynolds numbers of $1.6 \times 10^6$, $1.7 \times 10^6$, and $1.8 \times 10^6$ were intentionally measured using both instruments. For each velocity range, measurements adhered to the same procedure as in the first phase of experiments: the flow velocity was swept from a low to a high value and then back from high to low, with each complete cycle constituting one `run.'

\begin{figure}[H]
\centering
\includegraphics[width=1.0\textwidth]{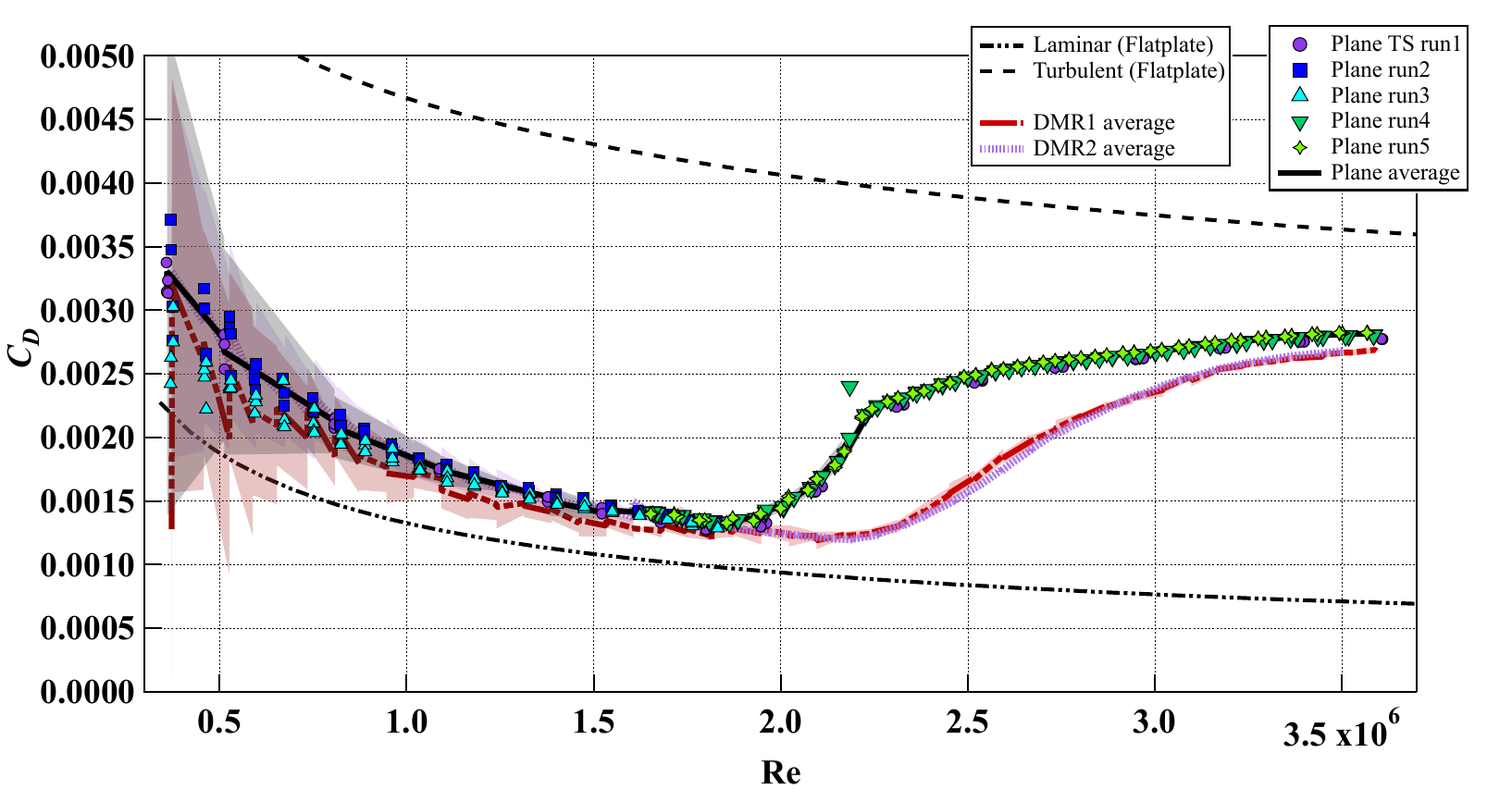}
\caption{Comparison of the total drag coefficient ($C_D$) versus the Reynolds number ($\mathit{Re}$) for the smooth (Plain), DMR1, and DMR2 surfaces in Phase II with tripping tapes applied. The dashed lines represent the theoretical skin friction coefficient ($C_f$) for laminar and turbulent flow (e.g., flat plate formulae). For the MSBS Experimental Data of the Smooth Surface (Plain), small circular ($\circ$), square ($\square$), triangle ($\triangle$), inverted-triangle ($\nabla$), and four-pointed star symbols show individual runs (Runs 1 to 5), and the thick solid line represents the averaged total drag. For the DMR1 and DMR2 Surfaces, the thick long-dotted line and the thick short-dotted line represent the averaged total drag for the respective surfaces.}
\label{fig:Plane}
\end{figure}

\begin{figure}[H]
\centering
\includegraphics[width=1.0\textwidth]{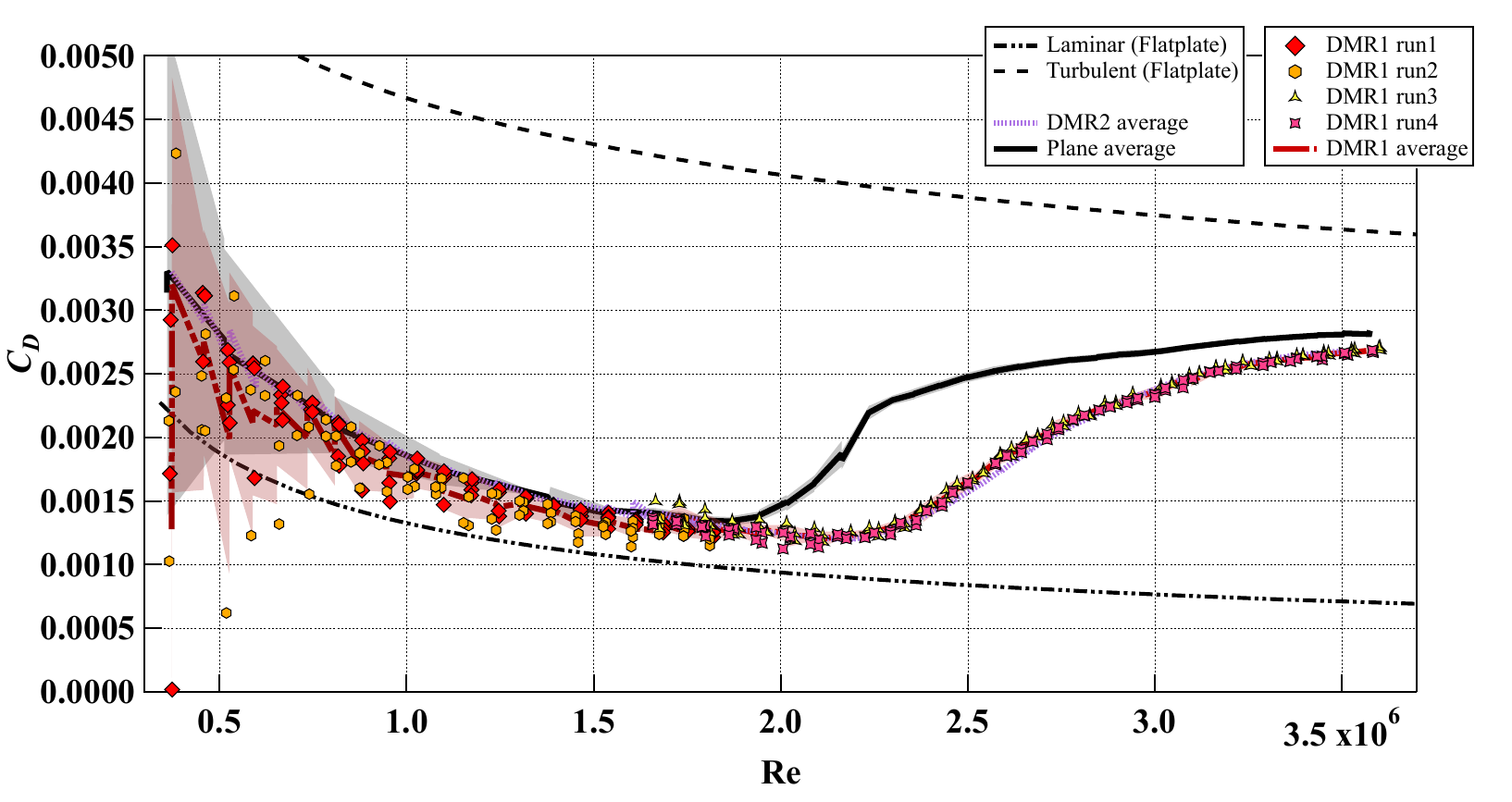}
\caption{Comparison of the total drag coefficient ($C_D$) versus the Reynolds number ($\mathit{Re}$) for the smooth (Plain), DMR1, and DMR2 surfaces in Phase II with tripping tapes applied. The dashed lines represent the theoretical skin friction coefficient ($C_f$) for laminar and turbulent flow (e.g., flat plate formulae). For the MSBS Experimental Data, individual runs are shown only for the DMR1 Surface by diamond ($\diamond$), hexagon, three-pointed star, and four-pointed star symbols (Runs 1 to 4). The averaged total drag is represented by the following lines: the thick long-dotted line for DMR1, the thick short-dotted line for DMR2, and the thick solid line for the Smooth Surface (Plain).}
\label{fig:DMR1}
\end{figure}

\begin{figure}[H]
\centering
\includegraphics[width=1.0\textwidth]{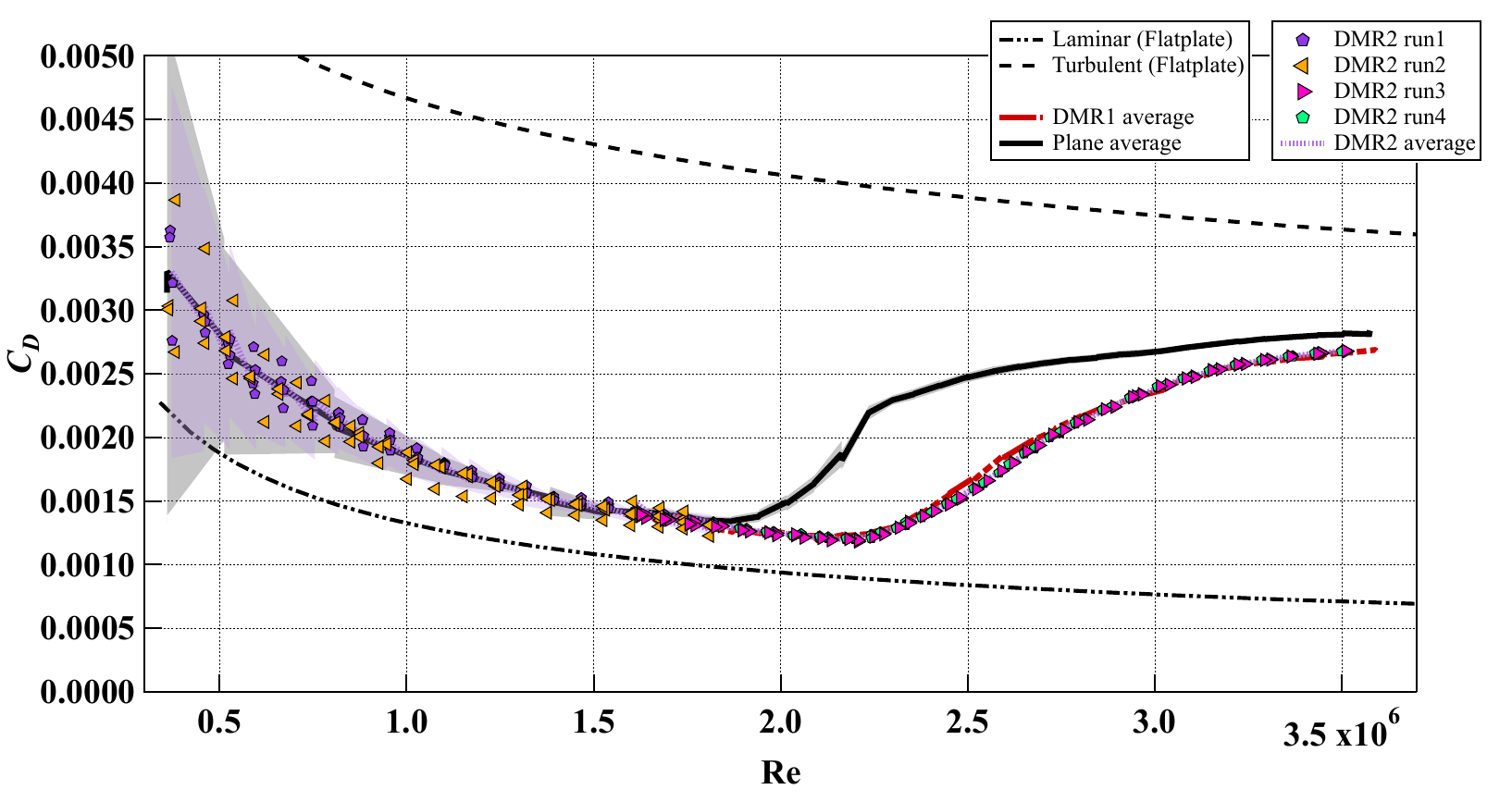}
\caption{Comparison of the total drag coefficient ($C_D$) versus the Reynolds number ($\mathit{Re}$) for the smooth (Plain), DMR1, and DMR2 surfaces in Phase II with tripping tapes applied. The dashed lines represent the theoretical skin friction coefficient ($C_f$) for laminar and turbulent flow (e.g., flat plate formulae). For the MSBS Experimental Data, individual runs are shown only for the DMR2 Surface by pentagon, left-pointing triangle ($\triangleleft$), right-pointing triangle ($\triangleright$), and light-coloured pentagon symbols (Runs 1 to 4). The averaged total drag is represented by the following lines: the thick solid line for the Smooth Surface (Plain), the thick long-dotted line for DMR1, and the thick short-dotted line for DMR2.}
\label{fig:DMR2}
\end{figure}

Both the raw and statistically processed measurement data are presented in our figures. To optimise the visual clarity and number of data points displayed per figure, Figure \ref{fig:Plane} illustrates the results from runs 1 to 5 for the smooth surface (Plain), alongside the statistically processed results for DMR1 and DMR2. Similarly, Figure \ref{fig:DMR1} includes the raw data from runs 1 to 4 for DMR1, in addition to the statistically processed results for the smooth surface and DMR2. Finally, Figure \ref{fig:DMR2} presents the raw data from runs 1 to 4 for DMR2, along with the statistically processed results for both the smooth surface and DMR1.

First, we confirmed that measurements from the two differential pressure gauges were consistent under identical Reynolds number conditions. Each measurement run followed the same path, exhibiting no hysteresis.
In this context, run 1 for the smooth surface in Figure \ref{fig:Plane} was conducted at the beginning of the experiment to provisionally investigate the transition Reynolds number. For this specific run, the entire Reynolds number range was measured using a single differential pressure gauge.

We observed a tendency for increased measurement error in the very low Reynolds number range, which was also noted in Phase I when only a single pressure gauge ($5\,\text{kPa}$ range) was used. This suggests that the flow field around this model exhibits significant fluctuations in the very low Reynolds number regime, potentially attributable to fluctuating separation and reattachment in the wake region. Notably, there was almost no difference in drag values between the smooth and DMR surfaces in the low Reynolds number range. For DMR1, we even observed a slight reduction in averaged drag compared to the smooth surface.

In experiments conducted during Phase II with Tripping Tape 2, as depicted in Figure \ref{fig:tape}(b), we observed that for the smooth surface, the critical Reynolds number at which transition began was approximately $1.9 \times 10^6$. Subsequently, as the Reynolds number increased, the drag coefficient ($C_D$) showed a parabolic increase, likely attributable to the linear amplification of turbulent energy due to the onset of Tollmien-Schlichting ($\text{TS}$) instability. This transition Reynolds number remained largely consistent across runs 1 to 5. On the other hand, for both DMR1 and DMR2, the critical Reynolds number was significantly shifted, increasing to approximately $2.2 \times 10^6$ for both cases. Similar to the smooth surface, DMR1 and DMR2 also followed nearly identical paths from run 1 to run 4, with no hysteresis observed. The maximum reduction in the drag coefficient ($C_D$) for DMR2 compared to the Plain surface occurred in the transition region. As a result, a maximum aerodynamic drag reduction of $43.6\%$ was achieved at approximately $Re = 2.25 \times 10^6$ in the transition region.

Remarkably, both DMR1 and DMR2 maintained lower drag coefficients compared to the smooth surface, extending up to the upper limit of the measured Reynolds numbers, approximately $3.6 \times 10^6$. This disparity in results likely stems from the differences in the geometrical characteristics between the glass-DMR used in Phase I and the DMR1 and DMR2 employed in Phase II. This finding suggests that optimising not only the roughness height and distribution but also the roughness shape could potentially yield a superior drag reduction effect.

The slight differences observed in the results between DMR1 and DMR2 may also be attributable to these variations in roughness shape. As mentioned in Section \ref{sec:dmr}, DMR2 featured fewer but deeper depressions compared to DMR1. Our Direct Numerical Simulations ($\text{DNS}$) have indicated that a reduced number of roughness elements (optimally around the boundary layer thickness) leads to a decrease in the increase of turbulent kinetic energy and a reduction in friction drag. In this experiment, DMR2 exhibited a slightly more reduced drag than DMR1, even when accounting for measurement errors. This suggests that parameters like the number of depressions (or, alternatively, the frequency of disturbances acting on the flow field) and the depression height (or depth), which are not captured by conventional roughness parameters such as the arithmetic mean roughness ($R_a$) or maximum height roughness ($R_y$), might be more effective in characterising the observed effects. We anticipate reporting on these detailed insights through ongoing parallel numerical simulations.

For clarity, a comprehensive summary of all test conditions, referred to as runs throughout this study, is compiled in Table \ref{tab:run_summary}.

\begin{table}
\centering
\begin{tabular}{lllll}
\toprule
\textbf{Test/Run Label} & \textbf{DMR Surface Type} & \textbf{Phase} & \textbf{Tripping Tapes} & \textbf{Pressure Gauge Range} \tabularnewline
\midrule
\addlinespace[1ex]
Plane (LES) & Smooth & N/A & No & N/A \tabularnewline
\midrule
\addlinespace[1ex]
\multicolumn{5}{l}{\textbf{Phase I Experiments (Figures 12 and 13)}} \tabularnewline
\midrule
\addlinespace[0.5ex]
Plane (Run 1, 2) & Smooth & I & No/Yes & $5\,\text{kPa}$ \tabularnewline
Plane (Average) & Smooth & I & No/Yes & $5\,\text{kPa}$ \tabularnewline
glass-DMR (Run 1, 2) & Glass Beads & I & No/Yes & $5\,\text{kPa}$ \tabularnewline
glass-DMR (Average) & Glass Beads & I & No/Yes & $5\,\text{kPa}$ \tabularnewline
\midrule
\addlinespace[1ex]
\multicolumn{5}{l}{\textbf{Phase II Experiments (Figures 14, 15, and 16)}} \tabularnewline
\midrule
\addlinespace[0.5ex]
Plane (Run 1) & Smooth & II & Yes & $5\,\text{kPa}$ \tabularnewline
Plane (Run 2, 3) & Smooth & II & Yes & $500\,\text{Pa}$ \tabularnewline
Plane (Run 4, 5) & Smooth & II & Yes & $5\,\text{kPa}$ \tabularnewline
Plane (Average) & Smooth & II & Yes & $500\,\text{Pa} / 5\,\text{kPa}$ \tabularnewline
DMR1 (Run 1, 2) & O-Well Inc. Type 1 & II & Yes & $500\,\text{Pa}$ \tabularnewline
DMR1 (Run 3, 4) & O-Well Inc. Type 1 & II & Yes & $5\,\text{kPa}$ \tabularnewline
DMR1 (Average) & O-Well Inc. Type 1 & II & Yes & $500\,\text{Pa} / 5\,\text{kPa}$ \tabularnewline
DMR2 (Run 1, 2) & O-Well Inc. Type 2 & II & Yes & $500\,\text{Pa}$ \tabularnewline
DMR2 (Run 3, 4) & O-Well Inc. Type 2 & II & Yes & $5\,\text{kPa}$ \tabularnewline
DMR2 (Average) & O-Well Inc. Type 2 & II & Yes & $500\,\text{Pa} / 5\,\text{kPa}$ \tabularnewline
\bottomrule
\end{tabular}
\caption{Summary of experimental and computational cases (runs) presented in Figures 12--16.}
\label{tab:run_summary}
\end{table}

\subsection{Oil Flow Visualisation for Cases with Tripping Tape 2 (Phase II)}

Oil flow visualisation experiments were also conducted to examine surface flow patterns on the model. The experimental setup is detailed below.

The oil mixture comprised Shin-Etsu Chemical Co. Ltd., KF96-50CS silicone oil (viscosity: $50\,\text{cSt}$) and Sinloihi FZ-2002 (Green) fluorescent pigment. Preparation involved adding $1\,\text{gram}$ of pigment to $100\,\text{ml}$ of silicone oil, followed by stirring with a magnetic stirrer for approximately one hour. Prior to application, the mixture was manually stirred again to ensure homogeneity. The oil was uniformly applied to the upper half of the model's trailing edge section by brushing. Image acquisition was performed using a NIKON D500 camera ($2848 \times 4272\,\text{pixels}$, $300\,\text{dpi}$, $24\text{-bit}$ colour depth) equipped with a $90\,\text{mm}$ focal length lens. Images were captured at $2\text{-second}$ intervals for approximately $15\,\text{minutes}$, commencing with the initiation of airflow. The camera was positioned above and downstream of the measurement section. For excitation, an HARDsoft IL-106X-UV LED light source ($390\,\text{nm}$ wavelength) with a $3\,\text{A}$ current was employed, positioned laterally and downstream of the measurement section.

\begin{figure}[H]
\centering
\subfigure[Plane Case]{
\includegraphics[width=0.4\textwidth]{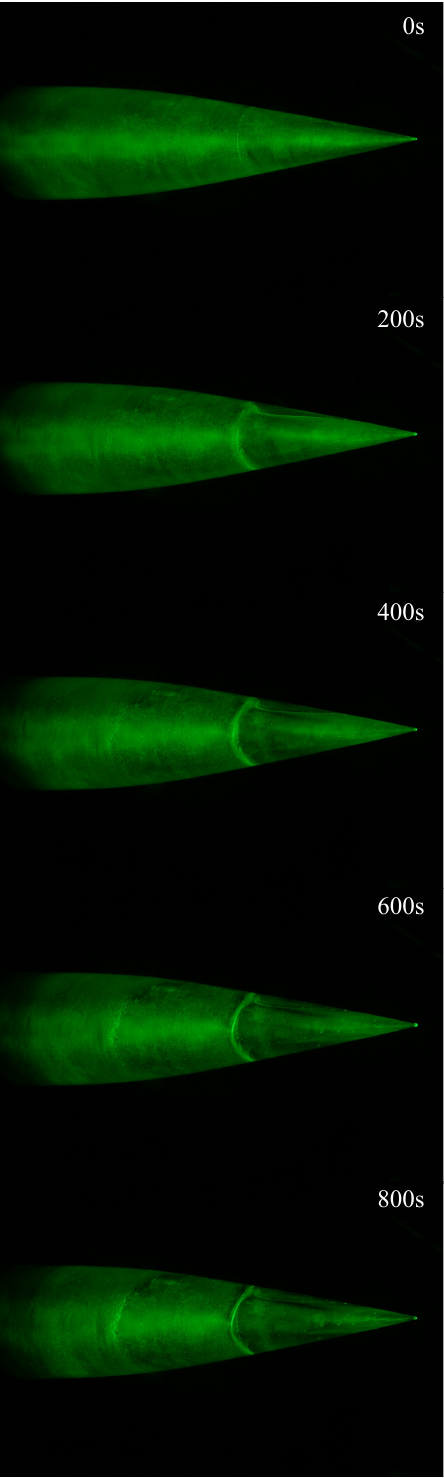}}
\hspace{3em}
\subfigure[glass-DMR case]{
\includegraphics[width=0.4\textwidth]{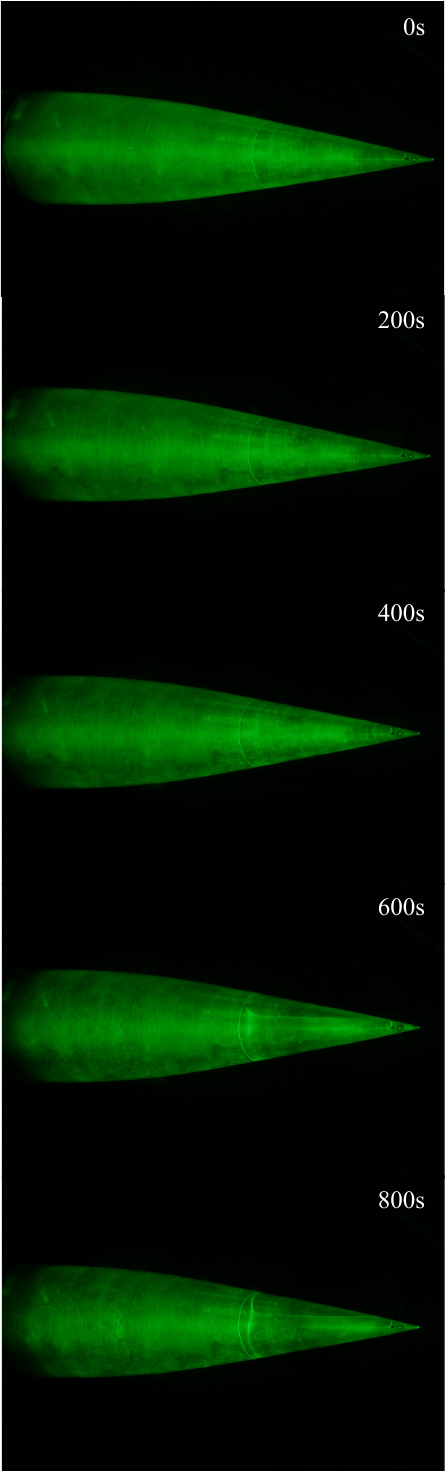}}
\caption{Oil flow visualisation for the smooth surface case and glass-DMR case at $Re = 1.2\times10^6$ (Phase II). Note that the measured total drag coefficient ($C_D$) is identical for both cases at this Reynolds number, despite the presence of localized oil accumulation near the tail.}
\label{fig:oilflow_low}
\end{figure}

Experiments were performed on a smooth surface model fitted with Tripping Tape 2 during Phase II. Mainstream velocities were set at two conditions: $17\,\text{m/s}$ (pre-transition, $Re \approx 1.2 \times 10^6$) and $47\,\text{m/s}$ (post-transition, $Re \approx 3.4 \times 10^6$). The resulting flow visualisations at each low and high Reynolds number for the smooth and glass-DMR surfaces are presented in Figures \ref{fig:oilflow_low}(a,b) and \ref{fig:oilflow_high}(a,b), respectively.

\begin{figure}[H]
\centering
\subfigure[Plane Case]{
\includegraphics[width=0.4\textwidth]{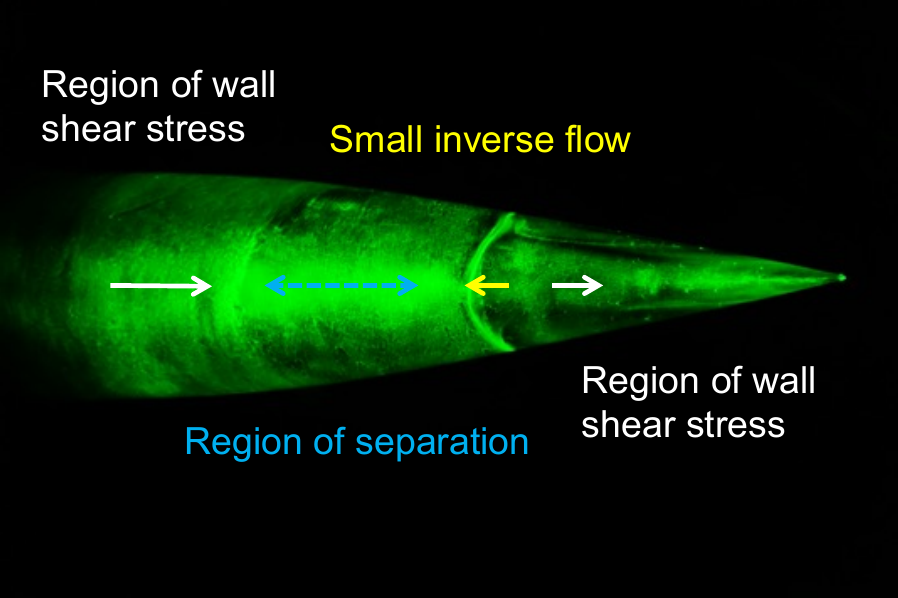}}
\hspace{3em}
\subfigure[glass-DMR case]{
\includegraphics[width=0.4\textwidth]{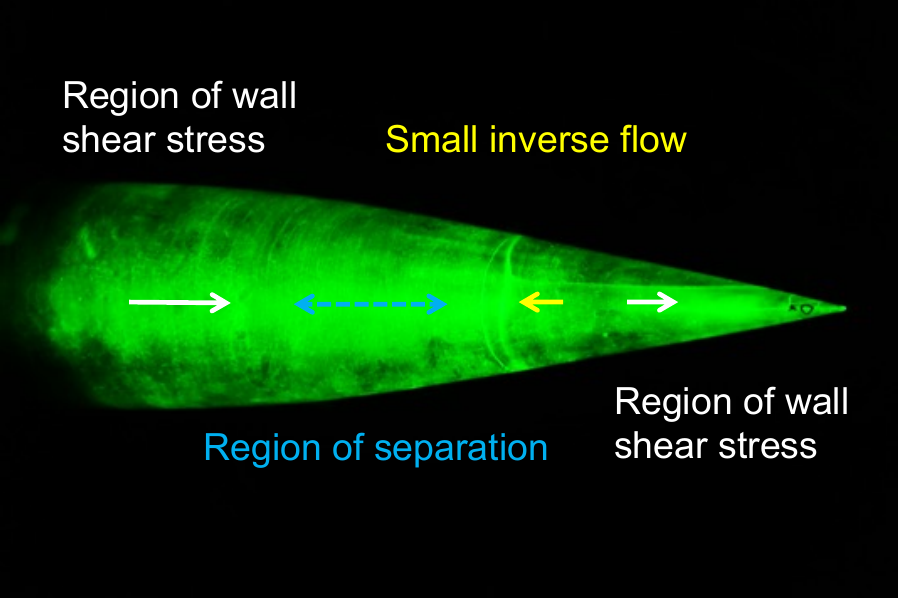}}
\caption{Oil flow visualisation with enhanced contrast and annotations of flow features for the smooth surface and glass-DMR cases at $Re = 1.2\times10^6$ (Phase II). Although localized separation regions and small inverse flows are identified, the consistency in the measured $C_D$ values  confirms that these structures do not contribute to a detectable change in total aerodynamic drag. This suggests that the observed oil accumulation does not represent a dominant pressure-drag mechanism, and the glass-DMR coating does not significantly influence the pressure-drag component in this regime.}
\label{fig:oilflow_enlarged}
\end{figure}

\begin{figure}[H]
\centering
\subfigure[Plane Case]{
\includegraphics[width=0.4\textwidth]{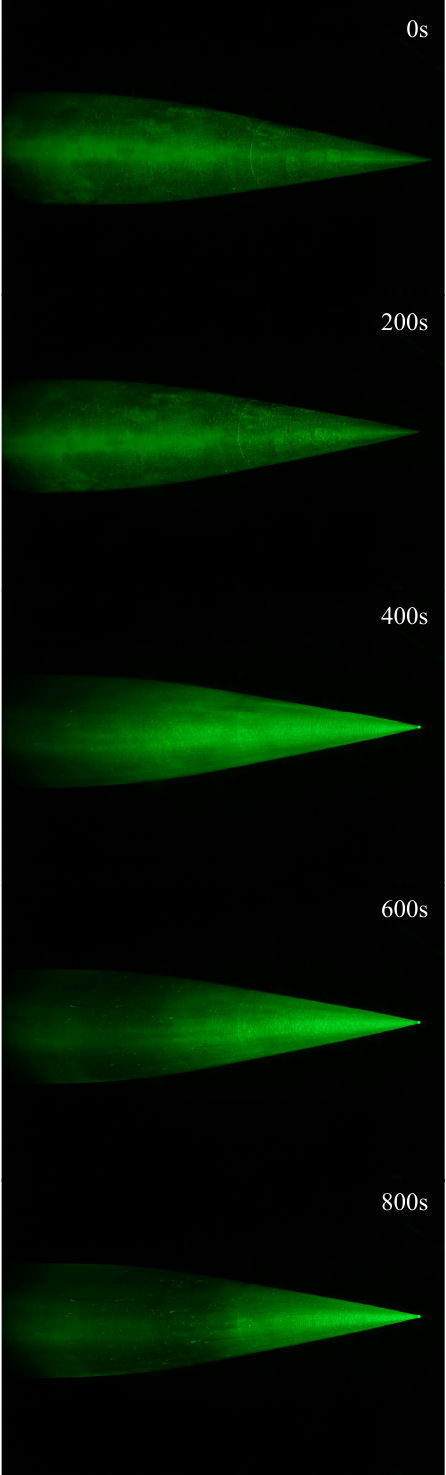}}
\hspace{3em}
\subfigure[glass-DMR case]{
\includegraphics[width=0.4\textwidth]{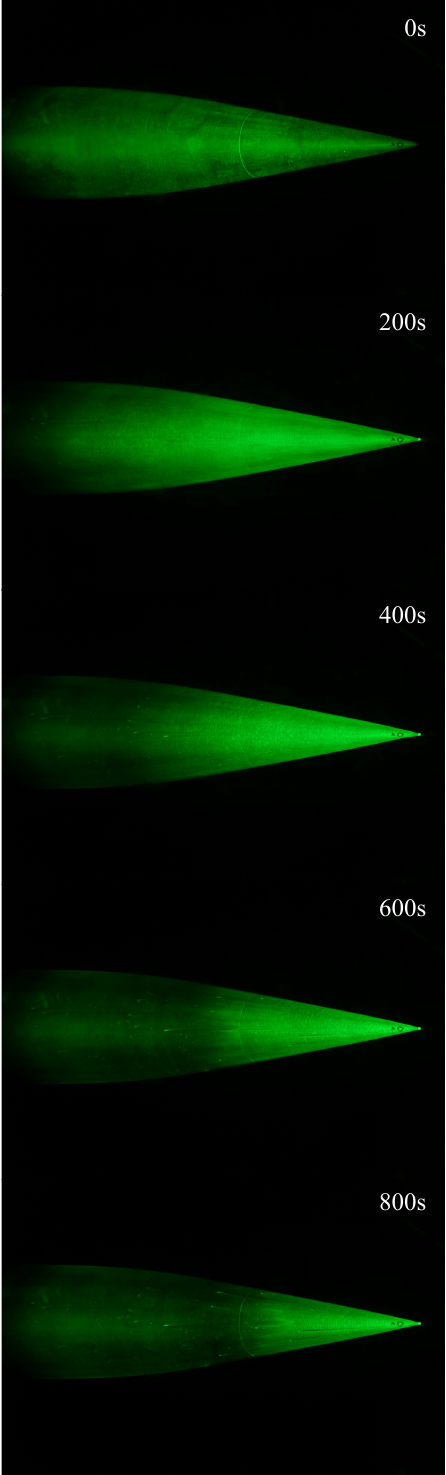}}
\caption{Oil flow visualisation for the smooth surface case and glass-DMR case at $Re = 3.4\times10^6$ (Phase II). At this higher Reynolds number, the oil is smoothly advected downstream without localized stagnation for both surfaces. The fact that a significant reduction in $C_D$ is observed for the glass-DMR surface in this attached-flow regime reinforces the conclusion that the drag benefit is independent of tail separation topology.}
\label{fig:oilflow_high}
\end{figure}

Figures \ref{fig:oilflow_low}(a) and (b) illustrate that at low Reynolds numbers, the flow field exhibited localized oil accumulation near the tail. As shown in the enhanced and annotated views (Figure \ref{fig:oilflow_enlarged}), these regions are characterized by localized separation and small inverse flows. Importantly, despite the presence of these separation patterns, the measured $C_D$ values for the smooth and glass-DMR surfaces remained identical at this Reynolds number (see Section \ref{sec:MSBS_PhaseII}). This observation provides critical physical evidence that these localized structures do not constitute a dominant drag component and that the glass-DMR coating does not significantly influence the pressure-drag component in this regime.

Conversely, figures \ref{fig:oilflow_high}(a) and (b) demonstrate that at high Reynolds numbers, the flow patterns sweep cleanly toward the trailing edge. Specifically, the oil streaks on the aft section show parallel trajectories for both the smooth and DMR cases at high Reynolds numbers, providing clear evidence of predominantly attached boundary layer flow. Notably, for the glass-DMR surface, the boundary layer remained attached even though its drag coefficient was higher than that of the smooth surface in this regime. This observation proves that the modification of aerodynamic drag—whether an increase or a decrease—is independent of the tail separation topology and confirms that the DMR effect is not realised through the suppression of tail separation, as no such separation exists to be controlled at these velocities.

The quantitative insignificance of separation suppression is further supported by the refined LES results discussed in \S \ref{sec:les}. Although the LES flow field remains predominantly laminar, the computed total pressure drag ($C_p \approx 0.00021$ at $Re = 3.6 \times 10^6$) provides a conservative upper bound for the pressure drag. In general, a turbulent boundary layer is more resistant to separation than a laminar one due to enhanced momentum exchange; therefore, the actual pressure drag in the experimental (turbulent) regime is expected to be even lower than, or at most comparable to, this laminar estimate. Since the measured drag reduction ($\Delta C_D \approx 0.001$) is nearly five times larger than this available $C_p$ budget, any potential modification of the tail flow by the DMR coating is numerically insufficient to explain the experimental benefit. This confirms that the observed reduction must be ascribed to changes in the skin friction coefficient ($C_f$) rather than separation control.

The Reynolds-number dependency of these oil-flow dynamics, observed regardless of the presence of the DMR coating, is shown in supplementary movies 1, 2, 3, and 4. At the lower Reynolds number ($Re = 1.2 \times 10^6$), the video reveals that a portion of the oil remains localised near the tail due to the small inverse flow for both the smooth and DMR surfaces. However, as noted previously, the measured $C_D$ values for both cases are identical at this Reynolds number, confirming that this localised stagnation has a negligible impact on the total aerodynamic drag. At higher Reynolds number, the video demonstrates that the oil is smoothly advected downstream without such stagnation on both the smooth and DMR models. This transition from localised stagnation to smooth advection, which is common to both surfaces, provides dynamic evidence that the flow is not dominated by separation at higher velocities. These results further support the conclusion that the measured drag reduction originates from the modification of the boundary layer state by the DMR rather than the suppression of a large-scale separation.

%============================================================================
% 								Conclusion
%============================================================================
\section{Conclusion\label{sec:conclusion}}

This study experimentally investigated the potential of distributed micro-roughness (DMR) for friction drag reduction on a streamlined body by influencing the boundary layer characteristics. Utilising the highly precise 1-m magnetic suspension and balance system (MSBS) at Tohoku University, we conducted comprehensive aerodynamic drag measurements, complemented by wall-resolved large-eddy simulations (LES) and dynamic oil-flow visualisation. The exceptional stability and accuracy of the MSBS proved crucial for discerning the subtle drag changes associated with the applied surface modifications.

Our initial experiments (Phase I) without artificial disturbances revealed that a glass-DMR coating generally caused a decrease in the critical Reynolds number due to promoted transition, consistent with the conventional understanding of surface roughness as a tripping mechanism. Wall-resolved LES results demonstrated excellent agreement with the theoretical laminar friction coefficient. Crucially, the LES decomposition revealed that the total pressure drag coefficient ($C_p$) is remarkably small, providing a quantitative "drag budget" of approximately $0.00021$ at $Re = 3.6 \times 10^6$. Since laminar boundary layers are generally more susceptible to separation than turbulent ones, this estimate provides a conservative upper bound for the pressure drag at this velocity. This confirms that the pressure drag contribution from the tail section is subordinate even under the most separation-prone conditions, establishing that the observed drag changes must primarily originate from the friction drag component.

A pivotal finding emerged from experiments incorporating artificial tripping tapes (Phases I and II). The application of DMR surfaces (glass-DMR, DMR1 and DMR2) increased the critical Reynolds number for drag rise. In Phase II with Tripping Tape 2, the critical Reynolds number for the smooth surface was approximately $1.9 \times 10^6$, whereas for the DMR surfaces (DMR1 and DMR2), it was significantly elevated to approximately $2.2 \times 10^6$. This increase directly translated into a maximum aerodynamic drag reduction of $43.6\%$ within the transitional Reynolds number range.

Furthermore, both DMR1 and DMR2 maintained lower drag coefficients than the smooth surface up to the highest measured Reynolds number of approximately $3.6 \times 10^6$. This sustained drag reduction indicates that specific geometrical characteristics of the DMR, beyond conventional roughness metrics, play a critical role in its efficacy. The subtle performance differences observed between DMR1 and DMR2---where DMR2 features fewer but deeper depressions---correlate with our previous direct numerical simulation (DNS) findings for flat-plate flows. Those simulations indicated that optimised roughness distributions can mitigate the intensity of turbulent energy and, consequently, reduce friction drag. These results underscore that parameters such as the spatial frequency and depth of surface features, which are not adequately captured by standard metrics like $R_a$ or $R_y$, are essential for characterising the observed aerodynamic benefits.

Finally, the hypothesis that the observed drag reduction results from suppressed flow separation (pressure drag reduction) is refuted by integrated experimental and numerical evidence. At low Reynolds numbers ($Re = 1.2 \times 10^6$), where separation is most likely to occur, oil-flow visualisation and supplementary video evidence revealed localised oil accumulation due to a small inverse flow for both surfaces. However, as noted previously, the measured $C_D$ values remained identical between the smooth and DMR cases. This confirms that the localised separation is not influenced by the DMR or, even should any subtle modification exist, it is insufficient to induce a detectable change in the total aerodynamic drag. At the higher Reynolds number investigated ($Re = 3.4 \times 10^6$), the oil was smoothly advected downstream without localised stagnation for both the smooth and DMR surfaces. Notably, for the glass-DMR surface, the boundary layer remained predominantly attached even though the drag coefficient was significantly higher than that of the smooth surface in this regime. This proves that the modification of aerodynamic drag—whether an increase or a decrease—is independent of the tail separation topology and that no discernible modification of the flow structure was induced by the DMR coating. Furthermore, since the total available pressure drag budget identified by LES is significantly smaller than the measured experimental drag reduction ($\Delta C_D \approx 0.001$), any potential modification of the pressure-drag component is both physically unobserved and numerically insufficient to account for the results.

In summary, this study conclusively demonstrates that optimised DMR can effectively achieve significant drag reduction on a streamlined body. The evidence confirms that the benefit is not driven by separation suppression but is ascribed to the modification of the boundary layer state and the resulting change in the skin friction coefficient ($C_f$). These findings challenge conventional wisdom regarding the sole role of roughness as a transition promoter and highlight the importance of surface geometry optimisation for passive flow control.

\begin{acknowledgments}
We extend our sincere gratitude to Mr. Kazuki Aoyama of O-Well Corporation for providing the DMR coating. We are also deeply thankful to Dr. Takuto Ogawa, Dr. Yasufumi Konishi, Mr. Naoki So, Mr. Muku Miyagi, Mr. Ryusei Haga, Ms. Kana Arima, Mr. Ryo Yoshioka and Prof. Akira Oyama for their invaluable assistance during the experiments.
The experiments were conducted utilising the 1-m Magnetic Suspension and Balance System (1-m MSBS) at the Low Turbulence Wind Tunnel Facility / Shock Wave Research Facilities of the Advanced Flow Experimental Research Center, Institute of Fluid Science, Tohoku University. The numerical simulations were performed on the "AFI-NITY II" supercomputer system, also located at the Advanced Fluid Information Research Center, Institute of Fluid Science, Tohoku University. We gratefully acknowledge the use of these facilities.
This study was financially supported by KAKENHI for Early-Career Scientists (Grant Nos. 19K14880 and 23K26030) from the Japan Society for the Promotion of Science (JSPS) under the Ministry of Education, Culture, Sports, Science and Technology (MEXT) of Japan. Further support was provided by the JST Forest Program (JPMJFR222R) and JST CREST (JPMJCR24Q6).
\end{acknowledgments}

\section*{Supplementary material}
Supplementary material is available at https://doi.org/10.1017/jfm.202X.XXX.
\begin{itemize}
\item Movie 1: High-speed oil-flow visualisation over the smooth surface at $Re = 3.4 \times 10^6$, showing a predominantly attached boundary layer.
\item Movie 2: High-speed oil-flow visualisation over the glass-DMR surface at $Re = 3.4 \times 10^6$. Despite a drag increase compared to the smooth surface due to the promotion of boundary layer transition at this Reynolds number, the flow remains predominantly attached. This confirms that the modification in drag is independent of the tail separation topology.
\item Movie 3: Low-speed oil-flow visualisation over the smooth surface at $Re = 1.2 \times 10^6$, illustrating localised oil accumulation and small inverse flow near the tail.
\item Movie 4: Low-speed oil-flow visualisation over the glass-DMR surface at $Re = 1.2 \times 10^6$, showing flow features virtually identical to the smooth surface and confirming that the DMR does not affect flow feature in this regime.
\end{itemize}

\section*{Data Availability Statement}
The data that support the findings of this study are available within the article and from the corresponding author upon reasonable request.

\section*{Declaration of Interests}
The authors report no conflicts of interest.

\section*{AI Tools Declaration}
Generative AI tools (Gemini, Google) were used for language refinement, grammar checking, and structural advice during the preparation of the final draft.

\appendix
\section{1-m MSBS Facility Details \label{sec:appendix}}

\subsubsection{Control System}
\begin{figure}[H]
\centering
\subfigure[System\label{fig:system}]{
\includegraphics[width=0.7\textwidth]{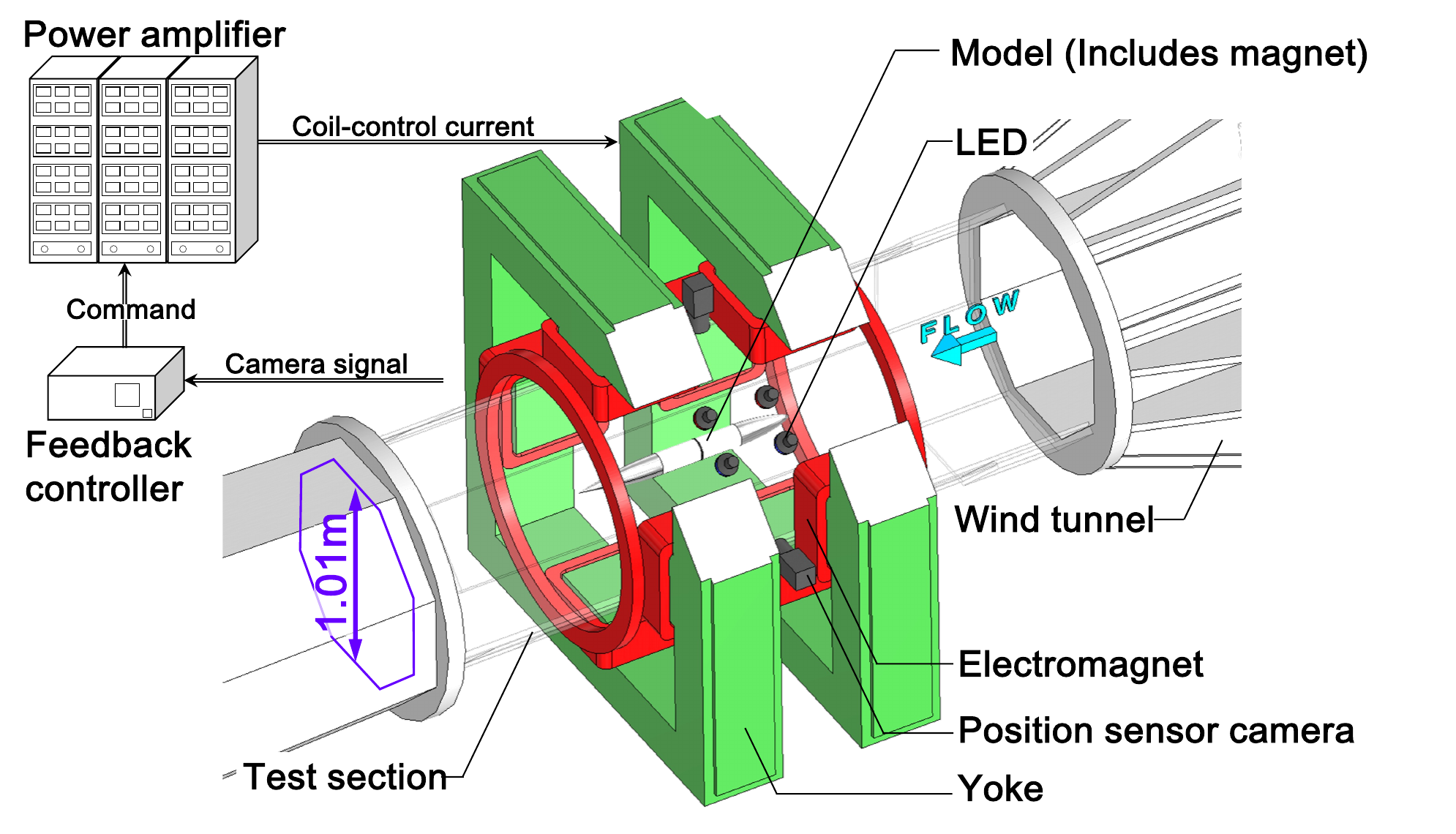}}
\par
\vspace{2em}
\subfigure[Control diagram \citep{okuizumi2018sports}.\label{fig:control}]{
\includegraphics[width=0.7\textwidth]{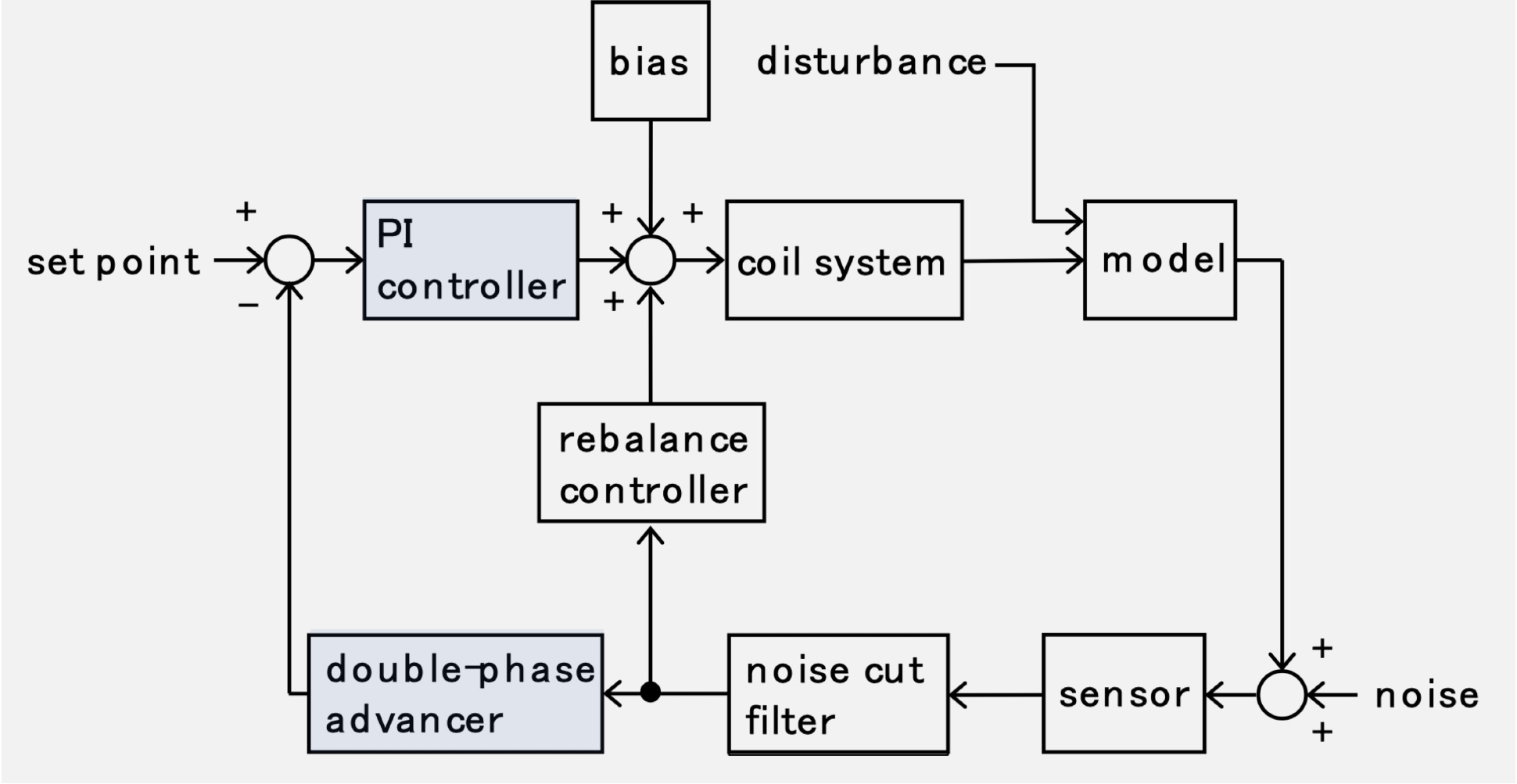}}
\caption{Control diagram of 1-m MSBS control system}
\end{figure}

The 1-m MSBS features an octagonal test section with a length of $3.0\,\text{m}$ and a width of $1.01\,\text{m}$, as depicted in Figure \ref{fig:msbs}. The control system is described in detail by \cite{okuizumi2018sports}. Ten coils are strategically arranged around the test section, each connected to a power amplifier capable of supplying up to $150\,\text{A}$ of current. To ensure efficient magnetic field generation given the large test section, a magnetic circuit is formed by connecting $8$ coils with a yoke, as shown in Figure \ref{fig:system}.

The position and attitude of the suspended model are precisely captured by five line sensor cameras, which utilise LED light sources and an advanced optical system. The readout frequency of these line sensor cameras is set at $1250\,\text{Hz}$. This high-speed, non-contact measurement of the model's position and attitude is achieved by appropriately reconfiguring the optical system based on the specific test model.

The stable suspension and precise positioning of the model are achieved through a unique Proportional-Integral ($\text{PI}$) control strategy, augmented by a double phase advancer, as illustrated in Figure \ref{fig:control}. This specific control architecture is vital for maintaining robust stability and dynamic response in the presence of aerodynamic disturbances, ensuring precise model attitude. The high sensor readout speed is critical for continuous and stable control, as any significant delay or infrequency in feedback could lead to a loss of control or instability, compromising the integrity of the experiment.

The 1-m MSBS employs a specialised position sensing system for accurate model control. Traditionally, the system has estimated the model's position by detecting its edges \citep{senda2018aerodynamic}, a method which proves effective for axisymmetric objects. For the present research, we applied a position measurement system that detects the model's edges via a distinctive black and white pattern to conduct our experiments, ensuring high-accuracy position and attitude measurement for our streamlined model. This edge-detection method, based on the contrast between the black and white painted sections of the model, is an established technique in the 1-m MSBS facility.

\subsubsection{Procedures}
Our experimental measurements commenced with the meticulous fabrication and preparation of the test model. As the MSBS suspends the model without physical contact, each model must be equipped with ferromagnetic cores that interact directly with the system's electromagnets. To ensure the highest accuracy in force and moment measurements and to guarantee stable suspension, the mass, centre of gravity, and magnetic properties of the model were meticulously measured and calibrated during this initial phase.

Following preparation, the model was carefully introduced into the wind tunnel's test section. The MSBS then engaged its sophisticated closed-loop control system, as previously outlined, to actively manage the electromagnetic forces. This system constantly monitors the model's precise position and attitude utilising optical sensors, dynamically adjusting the currents in the surrounding electromagnets. This enables the model to be stably suspended at its desired location and orientation throughout the experiment. Crucially, this control offers six degrees of freedom (three translations and three rotations), thereby allowing the precise adjustment of critical aerodynamic parameters such as angle of attack and yaw angle during testing. In the present series of tests, the levitated model was oriented in a fixed direction, parallel to the free stream flow, and a static measurement was conducted.

The 1-m MSBS facility is inherently capable of measuring all six force and moment components ($F_X$, $F_Y$, $F_Z$, and $M_X$, $M_Y$, $M_Z$). However, for the current axisymmetric model and experimental setup, which operates under five-degree-of-freedom control, the maximum measurable components are five (excluding the roll moment $M_X$). The five (or six) force and moment components acting on the model ($F_X$, $F_Y$, $F_Z$, and $M_X$, $M_Y$, $M_Z$) are derived from the measured currents of the ten electromagnets. This calculation utilises a set of simultaneous linear equations, with the coefficients (calibration matrix) determined beforehand through precise static force calibration procedures performed on the magnetic balance system \citep{okuizumi2018sports}. The drag force ($D$) is obtained directly as the $F_X$ component.

\bibliographystyle{jfm}
\bibliography{reference}

@article{Tani1935,
author = {Tani, I.},
title = {Effect of wing deflection surface on wing performance ({In Japanese})},
journal = {Mechanical Engineering Journal},
year = {1935},
pages = {743--744},
publisher = {The Japan Society of Mechanical Engineers}
}

@techreport{Tani1940,
author = {Tani, I. and Hama, R. and Mitsuisi, S.},
title = {On the permissible roughness in the laminar boundary layer ({In Japanese})},
institution = {Report of Aeronautical Research Institute, Tokyo Imperial University},
year = {1940},
pages = {417--428}
}

@techreport{Saric1997,
author = {Saric, W. S. and Reibert, M. S. and Carrillo, R. B. J.},
title = {Distributed-roughness effects on stability and transition in swept-wing boundary layer},
institution = {Contractor Report, {Langley Research Center}},
year = {1997},
number = {19970011192}
}

@article{Saric1994,
author = {Saric, W. S.},
title = {Low-speed boundary-layer transition},
journal = {Annual Review of Fluid Mechanics},
volume = {26},
number = {1},
pages = {379--409},
year = {1994},
publisher = {Annual Reviews},
doi = {10.1146/annurev.fl.26.010194.002115}
}

@inproceedings{Morkovin1969,
  title={On the many faces of transition},
  author={Morkovin, Mark V},
  booktitle={Viscous Drag Reduction: Proceedings of the Symposium on Viscous Drag Reduction held at the LTV Research Center, Dallas, Texas, September 24 and 25, 1968},
  pages={1--31},
  year={1969},
  organization={Springer}
}

@article{Schaffarczyk2017,
title={Experimental detection of laminar-turbulent transition on a rotating wind turbine blade in the free atmosphere},
author={Schaffarczyk, A. P. and Schwab, D. and Breuer, M.},
journal={Wind Energy},
volume={20},
number={2},
pages={211--220},
year={2017},
publisher={Wiley Online Library}
}

@article{Yakeno2021,
author = {Yakeno, A. and Obayashi, S.},
title = {Propagation of stationary and traveling waves in a leading-edge boundary layer of a swept wing},
journal = {Physics of Fluids},
volume = {33},
number = {9},
year = {2021},
pages = {094111},
doi = {10.1063/5.0063936}
}

@article{nakagawa2023effects,
title = {Effects of freestream turbulence on the secondary instability of the roughness-induced crossflow vortex in swept flat plate boundary layers},
author = {Nakagawa, K. and Ishida, T. and Tsukahara, T.},
journal = {International Journal of Heat and Fluid Flow},
volume = {102},
pages = {109161},
year = {2023},
publisher = {Elsevier}
}

@inproceedings{Mori2024,
author = {Mori, Y. and Yakeno, A. and Obayashi, S.},
title = {Effects of surface roughness and free-stream turbulence on transitions in the swept-wing boundary layer},
booktitle = {{AIAA SciTech Forum 2024}},
pages = {1565},
year = {2024},
}

@inproceedings{Mori2024IUTAM,
author = {Mori, Y. and Yakeno, A. and Ogawa, T. and Obayashi, S.},
title = {Numerical simulation of transition over a transonic swept wing with distributed roughness},
booktitle = {{IUTAM Symposium on Laminar-Turbulent Transition}},
address = {Nagano, Japan},
year = {2024}
}

@inproceedings{Mori2024AIAA,
author = {Mori, Y. and Yakeno, A. and Obayashi, S.},
title = {{DNS} of boundary-layer transition over a transonic swept wing under real flight condition},
booktitle = {{AIAA AVIATION FORUM AND ASCEND 2024}},
pages = {4486},
year = {2024}
}

@misc{AirbusBLADE,
title = {{Airbus}' “{BLADE}” laminar flow wing demonstrator makes first flight},
author = {Airbus},
howpublished = {\url{https://www.airbus.com/en/newsroom/press-releases/2017-09-airbus-blade-laminar-flow\\-wing-demonstrator-makes-first-flight}},
urldate = {2025-06-26},
year = {2017},
note = {Accessed on June 26, 2025}
}

@article{hamada2023drag,
title = {Drag reduction effect of distributed roughness on the transitional flow state using {Direct Numerical Simulation}},
author = {Hamada, S. and Yakeno, A. and Obayashi, S.},
journal = {International Journal of Heat and Fluid Flow},
volume = {104},
pages = {109230},
year = {2023},
publisher = {Elsevier}
}

@misc{patent7609489,
author = {Yakeno, A.},
title = {Evaluation apparatus, rough surface, evaluation method, and program},
year = {2024},
month = {December},
note = {Patent no. 7609489, Filed in Japan},
howpublished = {Japanese Patent}
}

@inproceedings{ogawanumerical,
title = {Numerical investigation of distributed roughness effects for transient flow},
author = {Ogawa, T. and Yakeno, A.},
booktitle = {Proceedings of 13th International Symposium on Turbulence and Shear Flow Phenomena ({TSFP13})},
year = {2024}
}

@inproceedings{ogawa2024parametric,
author = {Ogawa, T. and Yakeno, A.},
title = {Parametric study of distributed roughness effects for transitional flow},
booktitle = {Proceedings of the {International Conference on Computational Fluid Dynamics} ({ICCFD})},
year = {2024},
address = {Kobe}
}

@article{Rasheed2002Experiments,
title = {Experiments on passive hypervelocity boundary-layer control using an ultrasonically absorptive surface},
author = {Rasheed, A. and Hornung, H. G. and Fedorov, A. V. and Malmuth, N. D.},
journal = {{AIAA Journal}},
volume = {40},
pages = {481--489},
year = {2002},
publisher = {American Institute of Aeronautics and Astronautics},
doi = {10.2514/2.1691}
}

@article{Fedorov2003Stabilization,
title = {Stabilization of a hypersonic boundary layer using an ultrasonically absorptive coating},
author = {Fedorov, A. and Shiplyuk, A. and Maslov, A. and Burov, E. and Malmuth, N.},
journal = {Journal of Fluid Mechanics},
volume = {479},
pages = {99--124},
year = {2003},
publisher = {Cambridge University Press},
doi = {10.1017/S002211200200326X},
note = {Corrigendum in {J. Fluid Mech.} 2015, 769, 725--728}
}

@article{Lukashevich2018Passive,
title = {Passive porous coating effect on a hypersonic boundary layer on a sharp cone at small angle of attack},
author = {Lukashevich, S. and Morozov, S. and Shiplyuk, A.},
journal = {Experiments in Fluids},
volume = {59},
pages = {130},
year = {2018},
publisher = {Springer Berlin Heidelberg},
doi = {10.1007/s00348-018-2603-9}
}

@article{Lim2022Simulation,
title = {Simulation of hypersonic boundary layer on porous surfaces using {{OpenFOAM}}},
author = {Lim, J. and Kim, M. and Park, J. and Kim, T. and Jee, S. and Park, D.},
journal = {Computers \& Fluids},
volume = {240},
pages = {105437},
year = {2022},
publisher = {Elsevier}
}

@article{Lim2023Turbulent,
title = {Turbulent transition control using porous surfaces in hypersonic boundary layer},
author = {Lim, J. and Kim, M. and Bae, H. and Lin, R. S. and Jee, S.},
journal = {International Journal of Aeronautical and Space Sciences},
volume = {24},
number = {4},
pages = {972--984},
year = {2023},
publisher = {Springer}
}

@article{Running2023Attenuation,
title = {Attenuation of hypersonic second-mode boundary-layer instability with an ultrasonically absorptive silicon-carbide foam},
author = {Running, C. L. and Bemis, B. L. and Hill, J. L. and Borg, M. P. and Redmond, J. J. and Jantze, K. and Scalo, C.},
journal = {Experiments in Fluids},
volume = {64},
pages = {79},
year = {2023},
publisher = {Springer Berlin Heidelberg},
doi = {10.1007/s00348-023-03610-1}
}

@techreport{Holloway1964Effect,
title = {Effect of controlled surface roughness on boundary-layer transition and heat transfer at {Mach} numbers of 4.8 and 6.0},
author = {Holloway, P. F. and Sterrett, J. R.},
institution = {{NASA Langley Research Center}},
type = {Technical Report {NASA-TN-D-2054}},
address = {Hampton, {VA}, {USA}},
year = {1964}
}

@article{Duan2012Stabilization,
title = {Stabilization of a {Mach} 5.92 boundary layer by two-dimensional finite-height roughness},
author = {Duan, L. and Wang, X. and Zhong, X.},
journal = {{AIAA Journal}},
volume = {51},
pages = {266--270},
year = {2012},
publisher = {American Institute of Aeronautics and Astronautics},
doi = {10.2514/1.J051772}
}

@article{Fong2015Second,
title = {Second mode suppression in hypersonic boundary layer by roughness: {Design} and experiments},
author = {Fong, K. D. and Wang, X. and Huang, Y. and Zhong, X. and McKiernan, G. R. and Fisher, R. A. and Schneider, S. P.},
journal = {{AIAA Journal}},
volume = {53},
pages = {3138--3144},
year = {2015},
publisher = {American Institute of Aeronautics and Astronautics},
doi = {10.2514/1.J053896}
}

@inproceedings{Malmuth1998Problems,
title = {Problems in high speed flow prediction relevant to control},
author = {Malmuth, N. and Fedorov, A. and Shalaev, V. and Cole, J. and Hites, M. and Williams, D. and Khokhlov, A.},
booktitle = {2nd {AIAA}, {Theoretical Fluid Mechanics Meeting}},
pages = {2695},
year = {1998}
}

@article{Fedorov2001Stabilization,
title = {Stabilization of hypersonic boundary layers by porous coatings},
author = {Fedorov, A. V. and Malmuth, N. D. and Rasheed, A. and Hornung, H. G.},
journal = {{AIAA Journal}},
volume = {39},
pages = {605--610},
year = {2001},
publisher = {American Institute of Aeronautics and Astronautics},
doi = {10.2514/2.1384}
}

@article{Li2024Porous,
title = {Porous surface design with stability analysis for turbulent transition control in hypersonic boundary layer},
author = {Li, J. and Wang, Z. and Zhang, M. and Chen, Y.},
journal = {Applied Sciences},
volume = {12},
number = {6},
pages = {518},
year = {2024},
publisher = {{MDPI}},
url = {https://www.mdpi.com/2226-4310/12/6/518}
}

@article{Tani1989,
title = {Re-evaluation of {Nikuradse}'s experimental data for rough pipes},
author = {Tani, I.},
journal = {Proceedings of the Japan Academy, Series {B}},
volume = {65},
number = {6},
pages = {133-136},
year = {1989},
doi = {10.2183/pjab.65.133}
}

@article{Oguri1998,
title = {Control of turbulent transition using fiber surface in flat plate boundary layer ({In Japanese})},
author = {Oguri, E. and Kohama, Y.},
journal = {Transactions of the Japan Society of Mechanical Engineers. {B.}},
volume = {64},
number = {625},
pages = {2942-2949},
year = {1998},
doi = {10.1299/kikaib.64.2942}
}

@article{Oguri1996,
title = {Drag reduction by micro-sized distributed surface geometry on a flat plate ({In Japanese})},
author = {Oguri, E. and Kohama, Y.},
journal = {Transactions of the Japan Society of Mechanical Engineers. {B.}},
volume = {62},
number = {597},
pages = {1754-1761},
year = {1996},
doi = {10.1299/kikaib.62.1754}
}

@article{Kikuchi2004,
title = {Control of bypass transition for textile surface},
author = {Kikuchi, S. and Shimoji, M. and Watanabe, H. and Kohama, Y.},
journal = {{JSME International Journal Series B Fluids and Thermal Engineering}},
volume = {47},
number = {4},
pages = {777-785},
year = {2004},
doi = {10.1299/jsmeb.47.777}
}

@article{Nikuradse1933,
author = {Nikuradse, J.},
title = {Stromungsgesetze in rauhen rohren},
journal = {{VDI-Forschungsheft}},
volume = {361},
year = {1933}
}

@article{Schlichting1936,
author = {Schlichting, H.},
title = {Experimentelle untersuchungen zum rauhigkeitsproblem},
journal = {{Ingenieur-Archiv}},
volume = {7},
number = {1},
pages = {1--34},
year = {1936}
}

@article{Hama1954,
author = {Hama, F. R.},
title = {Boundary-layer characteristics for smooth and rough surfaces},
journal = {Transactions of the Society of Naval Architects and Marine Engineers},
volume = {62},
pages = {335--351},
year = {1954}
}

@article{Perry1969,
author = {Perry, A. E. and Schofield, W. H. and Joubert, P. N.},
title = {Roughness-induced skin friction in turbulent boundary layers},
journal = {Journal of Fluid Mechanics},
volume = {37},
number = {2},
pages = {383--401},
year = {1969}
}

@article{Flack2010,
author = {Flack, K. A. and Schultz, M. P.},
title = {The dependence of the roughness function on the {Reynolds} number and the roughness geometry},
journal = {Journal of Fluid Mechanics},
volume = {651},
pages = {89--122},
year = {2010}
}

@article{Bhaganagar2004,
author = {Bhaganagar, K. and Kim, J. and Coleman, G. N.},
title = {A numerical study of turbulent channel flow over a randomly rough surface},
journal = {Journal of Fluid Mechanics},
volume = {520},
pages = {243--271},
year = {2004}
}

@article{Busse2017,
author = {Busse, A. and Thakkar, M. and Sandham, N. D.},
title = {Direct numerical simulation of turbulent channel flow over isotropic roughness},
journal = {Journal of Fluid Mechanics},
volume = {829},
pages = {442--475},
year = {2017}
}

@article{nonomiya2024development,
title = {Development of a direct measurement device for the local wall shear stress in boundary layer flows},
author = {Nonomiya, T. and Sasamori, M. and Mochizuki, S.},
journal = {Journal of Fluid Science and Technology},
volume = {19},
number = {3},
pages = {{JFST0027}--{JFST0027}},
year = {2024},
publisher = {The Japan Society of Mechanical Engineers}
}

@book{butt2018transition,
title = {Transition to turbulence in low-dimensional systems: The {Butts} and {Egbers} diagrams},
author = {Butt, S. and Egbers, D.},
year = {2018},
publisher = {Springer International Publishing},
address = {Cham}
}

@article{squire2016comparison,
title = {Comparison of turbulent boundary layers over smooth and rough surfaces up to high {Reynolds} numbers},
author = {Squire, D. T. and Monty, M. J. and Marusic, I. and Hutchins, N.},
journal = {Journal of Fluid Mechanics},
volume = {796},
pages = {},
year = {2016},
publisher = {Cambridge University Press}
}

@article{chung2021predicting,
title = {Predicting the drag of rough surfaces},
author = {Chung, D. and Hutchins, N. and Schultz, M. P. and Flack, K. A.},
journal = {Annual Review of Fluid Mechanics},
volume = {53},
pages = {439--471},
year = {2021},
publisher = {Annual Reviews},
doi = {10.1146/annurev-fluid-062520-115127}
}

@techreport{tuttle1983magnetic,
title = {Magnetic suspension and balance systems: {A} selected, annotated bibliography},
author = {Tuttle, M. H. and Kilgore, R. A. and Boyden, R. P.},
institution = {{NASA Langley Research Center}},
type = {{NASA Technical Memorandum}},
number = {{NASA TM-84661}},
year = {1983}
}

@techreport{lawing1987magnetic,
title = {Potential benefits of magnetic suspension and balance systems},
author = {Lawing, P. L. and Dress, D. A. and Kilgore, R. A.},
institution = {{NASA Langley Research Center}},
type = {{NASA Technical Memorandum}},
number = {{NASA TM-89079}},
year = {1987}
}

@phdthesis{garbutt1992propulsion,
title = {Propulsion simulation in a magnetic suspension wind tunnel with special reference to force measurement},
author = {Garbutt, K. S.},
year = {1992},
school = {University of Southampton},
pages = {261},
address = {Southampton, {UK}}
}

@article{sawada1995recent,
title = {Recent progress of magnetic suspension and balance systems in {Japan}},
author = {Sawada, H.},
journal = {Journal of Aircraft},
volume = {32},
number = {4},
pages = {696--701},
year = {1995},
publisher = {American Institute of Aeronautics and Astronautics}
}

@article{okuizumi2024aerodynamic,
title = {Aerodynamic characteristics of magnetically suspended square cylinders with low fineness ratio},
author = {Okuizumi, H. and Makino, R. and Horiguchi, M. and Saito, Y. and Sawada, H. and Konishi, Y. and Obayashi, S. and Asai, K. and Nonomura, T.},
journal = {Journal of Aircraft},
pages = {1--7},
year = {2024},
publisher = {American Institute of Aeronautics and Astronautics}
}

@article{okuizumi2025wind,
title = {A wind tunnel test method for {Magnus} force measurement on rotating spheres using a 1-{M Magnetic Suspension and Balance System}},
author = {Okuizumi, H. and Sawada, H. and Konishi, Y. and Asai, K. and Nonomura, T. and Obayashi, S.},
journal = {Journal of Fluid Science and Technology},
volume = {20},
number = {1},
pages = {{JFST0002}},
year = {2025},
publisher = {The Japan Society of Mechanical Engineers}
}

@misc{JAL_JAXA_Owell_Nikon_2023,
author = {{JAL}},
title = {{JAL}, {JAXA}, {O-Well}, and {Nikon} conduct flight test with aircraft for which the world's first {Riblet} shape was applied over external paint},
howpublished = {\url{https://press.jal.co.jp/en/release/202303/007277.html}},
year = {2023},
month = {Mar},
note = {Accessed on 2025-07-03}
}

@misc{JAL_JAXA_Owell_2023,
author = {{JAL}},
title = {{JAL} initiates flight test to measure the fuel efficiency improvement effect by applying {Riblet} shapes in a large area on aircraft},
howpublished = {\url{https://press.jal.co.jp/en/release/202311/007781.html}},
year = {2023},
month = {Nov},
note = {Accessed on 2025-07-03}
}

@article{Ito_Kobayashi_Kohama_1992,
author = {Ito, H. and Kobayashi, R. and Kohama, Y.},
title = {The low-turbulence wind tunnel at {Tohoku University}},
volume = {96},
DOI = {10.1017/S0001924000024738},
number = {954},
journal = {The Aeronautical Journal},
year = {1992},
pages = {141--151}
}

@inproceedings{kohama1992tohoku,
title = {{Tohoku University Low-Turbulence Wind Tunnel}},
author = {Kohama, Y. and Kobayashi, R. and Ito, H.},
booktitle = {28th {Joint Propulsion Conference} and {Exhibit of AIAA}},
pages = {3913},
DOI = {10.2514/6.1992-3913},
year = {1992}
}

@article{kohama1982performance,
title = {Performance of small-scale low-turbulence wind tunnel},
author = {Kohama, Y. and Kobayashi, R. and Ito, H.},
journal = {The Memoirs of the Institute of High Speed Mechanics (Tohoku University)},
volume = {48},
pages = {119--142},
year = {1982},
note = {(In Japanese)}
}

@article{kohama1987some,
author = {Kohama, Y.},
title = {Some expectation on the mechanism of cross-flow instability in a swept wing flow},
journal = {Acta Mechanica},
volume = {66},
pages = {21--38},
year = {1987},
doi = {10.1007/BF01184283}
}

@inproceedings{kohama1999effective,
author = {Kohama, Y. and Egami, Y.},
title = {Effective laminar flow control by selective suction system on swept wing flow},
booktitle = {37th {Aerospace Sciences Meeting} and {Exhibit}},
year = {1999},
pages = {921},
doi = {10.2514/6.1999-921}
}

@article{kohama1994traveling,
author = {Kohama, Y. and Motegi, D.},
title = {Traveling disturbance appearing in boundary layer transition in a yawed cylinder},
journal = {Experimental Thermal and Fluid Science},
volume = {8},
number = {4},
pages = {273--278},
year = {1994},
doi = {10.1016/0894-1777(94)90057-4}
}

@inproceedings{suzuki2024experimental,
title = {Experimental validation of the suppression of crossflow instability by sinusoidal roughness elements},
author = {Suzuki, S. and Yakeno, A. and Konishi, Y. and Tokugawa, N. and Hirota, M. and Takami, H. and Obayashi, S.},
booktitle = {{AIAA SCITECH 2024 Forum}},
pages = {0891},
year = {2024}
}

@inproceedings{okuizumi2018sports,
author = {Okuizumi, H. and Sawada, H. and Nagaike, H. and Konishi, Y. and Obayashi, S.},
title = {Introduction of 1-{M} {MSBS} in {Tohoku University}, new device for aerodynamics measurements of the sports equipment},
booktitle = {Proceedings of the 12th {Conference of the International Sports Engineering Association}},
address = {Brisbane, Queensland, Australia},
month = {March},
year = {2018},
pages = {273},
series = {Proceedings},
volume = {2},
number = {6},
doi = {10.3390/proceedings2060273},
note = {Presented at the 12th Conference of the International Sports Engineering Association, {Brisbane}, {Queensland}, {Australia}, 26--29 {March} 2018. Published: 13 {February} 2018.}
}

@article{somers1980design,
author = {Somers, D. M.},
title = {Design and experimental results for the {{NLF-0415}} airfoil},
journal = {{NASA Technical Memorandum}},
year = {1980},
publisher = {{NASA}},
note = {{TM 81878}}
}

@inproceedings{senda2018aerodynamic,
author = {Senda, H. and Sawada, H. and Okuizumi, H. and Konishi, Y. and Obayashi, S.},
title = {Aerodynamic measurements of {{AGARD-B}} model at high angles of attack by 1-{M Magnetic Suspension and Balance System}},
booktitle = {{AIAA SciTech 2018 Forum}},
year = {2018},
note = {{AIAA Paper 2018-0302}},
publisher = {American Institute of Aeronautics and Astronautics},
doi = {10.2514/6.2018-0302}
}

@inproceedings{saric1991boundary,
author = {Saric, W. S. and Hoos, J. A. and Radeztsky, R. H.},
title = {Boundary-layer receptivity of sound with roughness},
booktitle = {{Boundary Layer Stability and Transition to Turbulence}; {Proceedings} of the {Symposium}, {ASME} and {JSME Joint Fluids Engineering Conference}, 1st},
address = {Portland, {OR}},
month = {June},
year = {1991},
pages = {17--22},
publisher = {American Society of Mechanical Engineers},
organization = {{ASME} and {JSME Joint Fluids Engineering Conference}, 1st}
}

@article{saric2002boundary,
author = {Saric, W. S. and Reed, H. L. and Kerschen, E. J.},
title = {Boundary-layer receptivity to freestream disturbances},
journal = {Annual Review of Fluid Mechanics},
volume = {34},
pages = {291--319},
year = {2002},
publisher = {Annual Reviews},
doi = {10.1146/annurev.fluid.34.091701.125944}
}

@misc{JP6511612B2,
author = {Asai, M. and Shinohara, R. and Nishizawa, H.},
title = {{Riblet Transfer Sheet}, {Riblet Transfer Sheet Manufacturing Method}, {Riblet Molding Method}, and {Transfer Sheet}},
year = {2019},
month = {May},
note = {Patent no. {JP6511612B2}, {Issued in Japan}. {Assignee}: {National Research and Development Agency Japan Aerospace Exploration Agency}, {O-Well Corp}, {Tokyo Metropolitan Univ.}},
howpublished = {Japanese Patent}
}

@book{Pankhurst1952,
  author    = {Pankhurst, R. C. and Holder, D. W.},
  title     = {Wind-tunnel technique},
  publisher = {Pitman Publishing},
  year      = {1952},
}

@techreport{joppa1973wind,
  title={Wind tunnel interference factors for high-lift wings in closed wind tunnels},
  author={Joppa, R. G.},
  year={1973},
  institution={NASA}
}

@book{britcher2023wind,
  title={Wind tunnel test techniques: design and use at low and high speeds with statistical engineering Applications},
  author={Britcher, C. and Landman, D.},
  year={2023},
  publisher={Academic Press}
}

@book{schlichting2017boundary,
  title={Boundary-Layer theory},
  author={Schlichting, H. and Gersten, K.},
  year={2017},
  edition={9},
  publisher={Springer},
  address={Berlin, Heidelberg}
}

@techreport{mack1984boundary,
  title={Boundary-layer linear stability theory},
  author={Mack, L. M.},
  year={1984},
  institution={Jet Propulsion Laboratory, California Institute of Technology},
  address={Pasadena, CA}
}

\end{document}